\documentclass[]{interact}

\usepackage{epstopdf}
\usepackage[caption=false]{subfig}

\usepackage[numbers,sort&compress]{natbib}
\usepackage{amsmath,amssymb}
\bibpunct[, ]{[}{]}{,}{n}{,}{,}
\makeatletter
\def\NAT@def@citea{\def\@citea{\NAT@separator}}
\makeatother

\theoremstyle{plain}

\theoremstyle{definition}

\theoremstyle{remark}

\begin{document}

\articletype{REGULAR ARTICLE}

\title{Orientational transition in nanobridges of nematic liquid crystals in slit pores}

\author{
\name{L.~F.~Rull\textsuperscript{a} and J.~M.~Romero-Enrique\textsuperscript{a,b}\thanks{CONTACT J.~M.~Romero-Enrique. Email: enrome@us.es}}
\affil{ \textsuperscript{a}Departamento de F\'{\i}sica At\'omica, Molecular y
Nuclear, \'Area de F\'{\i}sica Te\'orica, Universidad de Sevilla,
Avenida de Reina Mercedes s/n, 41012 Seville, Spain;  \textsuperscript{b}Instituto Carlos I de F\'{\i}sica Te\'orica y Computacional, Campus Universitario Fuentenueva,
Calle Dr. Severo Ochoa, 18071 Granada, Spain}
}

\maketitle

\begin{abstract}
In this work, the morphology of nematic capillary nanobridges in slit pores separated by a vertical distance $D$ will be characterised by Monte Carlo simulations for oblate molecules nematogen modeled by the Gay-Berne potential.
Previous studies on droplets show that the molecules are arranged homeotropically at the nematic-vapor interface and form spherical droplets with an annular disclination located in their equatorial plane. In the presence of pores with attractive substrates that favor homeotropic anchoring, the formation of nanobridges characterized by anchoring angles that decrease with increasing particle-substrate interaction intensity is observed. When the orientational field in the nanobridge is analyzed, the formation of an annular disclination of topological charge +1/2 located in the plane perpendicular to the z axis that passes through the center of mass of the nanobridge is observed for small $D$. However, when considering higher values of $D$, a change to a biaxial orientational profile within the nanobridge is observed, where now the annular disclination is arranged in a plane perpendicular to one direction of the $xy$ plane. These results are indicative of the existence of an orientational phase transition for an intermediate $D$ value between a uniaxial and a biaxial orientational configuration in the capillary nanobridge. 
\end{abstract}


\section{Introduction}

The long-range orientational order of nematic liquid crystals can be easily frustrated by the presence of boundary surfaces, such as walls or interfaces. The interplay between the elastic distortions, surface tension and anchoring allows for many different configurations in which the observation of topological defects is widespread \cite{Kleman1983}. The characterization of these states is interesting not only from a fundamental point of view, but because of their possible applications. As an example, the nematic droplets have been extensively studied in the last years \cite{Bernal1941111,Sonin1998129,Lev2001217061,Herring195187,Chandrasekhar1966,Williams1985,Williams1986,Virga1994,
Kaznacheev200257,Kaznacheev20031159,Prinsen2003,Prinsen200435}. They have been proposed as key ingredients 
for applications as privacy windows and other electro-optic devices \cite{Drzaic1995,Sutherland19941074,Bunning200083,Bowley20019,Jazbinsek20013831,Rudhardt20032610,Fernandez-Nieves2006}. 

In this paper we address the  configuration of capillary bridges confined in slit capillaries. The motivation for this work is a recent experimental and theoretical work \cite{Ellis2018} on 5CB capillary bridges on air in a slit pore formed by parallel glass surfaces with homeotropic anchoring conditions. This work reports different nematic configurations in the capillary bridge by tuning its diameter-to-height aspect ratio $\Gamma$ and shape, which can be either barrel-like or hourglass-like. So, for large $\Gamma$
 a ring defect is observed, while a point-defect structure is observed for small $\Gamma$. On the other hand, the defects are either hyperbolic for hourglass-like bridges or radial for barrel-like bridges. Our goal is to study this phenomenology via molecular simulations. For this purpose, we have considered the Gay-Berne model for nematogens \cite{Gay19813316} . This model has been used to model realistic liquid crystals. For example, 
the Gay-Berne model with $(\kappa,\kappa',\mu,\nu)=
(4.4,20,1,1)$ has been used to parametrize the calamitic $p$-terphenyl liquid crystal 
\cite{Luckhurst1993233,Bates19997087}, and
$(0.345, 0.2, 1, 2)$ for the triphenylene core discotic model \cite{Emerson1994113,Cienega-Cacerez20143171}. 
On the other hand, the phase diagram of the Gay-Berne model has been 
studied extensively
\cite{DeMiguel19901223,DeMiguel1991593,Chalam1991357,deMiguel1991174,deMiguel1991405,DeMiguel19923813}. 
For prolate molecules, the effect on the phase 
diagram of $\kappa$ and $\kappa'$ has been studied in Refs. 
\cite{Rull1995113,DeMiguel19964234,Brown19986685}.
Regarding oblate molecules, most of the simulation studies focus on the 
nematic-columnar phase transition
\cite{Emerson1994113,Bates19966696,Caprion2003417031,
Bellier-Castella20044847,Chakrabarti2008}. 

The paper is organized as follows. Section \ref{section2} is devoted the computer simulation methodology to generate capillary nanobridges. In Section \ref{section3} we describe the principal findings of our simulations. We end up the paper with a discussion of our results and main conclusions. 

\section{Methodology}
\label{section2}

\subsection{Potential model}
The Gay-Berne potential \cite{Gay19813316} describes the intermolecular interactions in nematogen fluids
by a suitable modification of the Lennard-Jones potential:
\begin{equation}
U_{ij} ({\bf r}_{ij},{\bf u}_i,{\bf u}_j)
= 4\varepsilon (\hat{\bf r}_{ij}, {\bf u}_i, {\bf u}_j)
\left[\rho_{ij}^{-12}-\rho_{ij}^{-6}\right]
\end{equation}
with
\begin{equation}
\rho_{ij} = \frac{ r_{ij} - \sigma (\hat{\bf r}_{ij},
{\bf u}_i, {\bf u}_j) + \sigma_0 }{\sigma_0}
\end{equation}
where $\bf u_{i}$ is the unit vector along the symmetry axis
of particle $i$, $r_{ij}=|{\bf r}_i - {\bf r}_j|$ is the distance
along the intermolecular vector $\bf{r}_{ij}$ joining the centers
of mass of particles $i$ and $j$,
and $\hat{\bf r}_{ij}={\bf r}_{ij}/r_{ij}$.
The anisotropic contact distance,
$\sigma(\hat{\bf r}_{ij},\bf{u}_i,\bf{u}_j)$, and the depth of the
interaction energy, $\varepsilon (\hat{\bf r}_{ij}, {\bf u}_i,
{\bf u}_j)$,
depend on the orientational unit vectors, the length-to-breath ratio
of the particle, $\kappa = \sigma_{ee} / \sigma_{ss}$, and the
energy depth anisotropy, $\kappa'= \epsilon_{ee} / \epsilon_{ss}$,
which are both defined as the ratio of the size and energy interaction
parameters in the end-to-end ($ee$) and side-by-side ($ss$)
configurations, respectively. With these definitions, $\kappa > 1$ 
corresponds to prolate particles while $\kappa < 1$ corresponds to oblate 
particles. Their expressions are given in terms
of an arbitrary length scale, $\sigma_0$, and an arbitrary energy
scale, $\epsilon_0$:
\begin{eqnarray}
\frac{\sigma (\hat{\bf r}_{ij}, {\bf u}_i, {\bf u}_j)}{\sigma_0}=
\Bigg[1-\frac{\chi}{2}\Bigg(\frac{(\hat{\bf r}_{ij}\cdot {\bf u}_i
+\hat{\bf r}_{ij}\cdot {\bf u}_j)^2}{1+\chi({\bf u}_i \cdot {\bf u}_j)}
+\frac{(\hat{\bf r}_{ij}\cdot {\bf u}_i
-\hat{\bf r}_{ij}\cdot {\bf u}_j)^2}{1-\chi({\bf u}_i \cdot {\bf u}_j)}
\Bigg)\Bigg]^{-1/2}
\end{eqnarray}
and
\begin{eqnarray}
\frac{\varepsilon (\hat{\bf r}_{ij}, {\bf u}_i,{\bf u}_j)}{\epsilon_0}=
[\epsilon_1({\bf u}_i,{\bf u}_j)]^{\nu}\times [\epsilon_2(\hat{\bf r}_{ij},
{\bf u}_i,{\bf u}_j)]^{\mu},
\end{eqnarray}
where
\begin{eqnarray}
\epsilon_1({\bf u}_i,{\bf u}_j)&=&
[1-\chi^2({\bf u}_i \cdot {\bf u}_j)^2]^{-1/2},
\\
\epsilon_2(\hat{\bf r}_{ij},{\bf u}_i,{\bf u}_j)&=&
1-\frac{\chi'}{2}\Bigg[\frac{(\hat{\bf r}_{ij}\cdot {\bf u}_i
+\hat{\bf r}_{ij}\cdot {\bf u}_j)^2}{1+\chi'({\bf u}_i \cdot {\bf u}_j)}
+\frac{(\hat{\bf r}_{ij}\cdot {\bf u}_i
-\hat{\bf r}_{ij}\cdot {\bf u}_j)^2}{1-\chi'({\bf u}_i \cdot {\bf u}_j)}
\Bigg],
\end{eqnarray}
$\chi = (\kappa^2 - 1) / (\kappa^2 +1)$ and $\chi'=[
(\kappa')^{1/\mu} - 1] / [(\kappa')^{1/\mu} + 1]$. 
Here $\sigma_0$ is the side-by-side 
intermolecular collision diameter, and $\epsilon_0$ is $[2\kappa/(\kappa^2+1)]^\nu$ times the minimum 
intermolecular potential energy between two molecules in the side-by-side 
configuration. As in the original paper by Gay and Berne  \cite{Gay19813316}, we choose $\mu = 2$ and $\nu=1$. 
With this election, the key parameters are the length-to-breadth geometrical ratio $\kappa$ and
$\kappa'$, which plays an important role in the formation of ordered
phases, as well as  the nematic anchoring on the nematic-vapor phase 
\cite{Mills19983284,Rull2017}. So, if $\kappa\le \kappa'$, the nematic
anchors homeotropically to the nematic-vapor interface, and  
random-planar otherwise. 

The interaction between the slit pore of width $D$ along the $z$ axis and the particle $i$ located at a height $z_i$ is given by
\begin{equation}
U_{w,i}(z_i)=U_{GB,w}(z_i)+U_{GB,w}(D-z_i),
\label{U_wall}
\end{equation}
where the interaction between the single wall and a particle $i$ is a $9-3$ potential:
\begin{equation}
U_{GB,w}(z)=a\frac{2\pi}{3}\rho_w\sigma_w^3 \epsilon_w\left[\frac{2}{15}\left(\frac{\sigma_w}z\right)^9-\left(\frac{\sigma_w}z\right)^3\right],
\label{U_wall_2}
\end{equation}
where $z$ is the distance between the wall and the center of mass of the particle, $\rho_w \sigma^3=0.988$, $\sigma_w/\sigma_0=1.096$, $\epsilon_w/\epsilon_0=1.277$ and $a$ is an adimensional wall interaction strength which can be varied. Note that, for $a=1$, $U_{GB,w}$ corresponds to the wall-particle interaction for the Ar-CO$_2$ system \cite{Finn1989}. On the other hand, although the potential is not dependent on the orientation of the GB particle (as it was considered in Ref. \cite{Rull2007}), we will see that this wall promotes a preferred anchoring to the GB particles. 

\subsection{Simulation procedure to generate the nanobridges}

The procedure to generate the nanobridges is analogous to the considered to generate nanodroplets in Refs. \cite{Rull2012,Rull2017}. As a first step, we 
obtain a near-coexistence bulk nematic phase by performing an isothermal-isobaric Monte Carlo simulation ($NPT-MC$) at zero pressure \cite{Frenkel1996}. We consider as simulation box a square cuboid in which the side length along the $z$ axis takes a fixed value $D$, but the side lengths along the $x$ and $y$ directions are allowed to fluctuate. 
Starting from a configuration of $N$ particles with higher density than the expected nematic-vapor coexisting value, the system decreases its density until it reaches the value of the nematic branch at coexistence. However, simulations are short to prevent large fluctuations which may lead the system to the vapor region. In addition, the maximum trial volume change is small, but tuned to get an acceptance ratio of about 30\% in volume change trial movements \cite{Rull2017}. 

Once the coexisting nematic phase is equilibrated, we use the last configuration of the $NPT-MC$ simulation as a seed for the simulation in the slit pore. For this purpose, the final configuration of the bulk simulation is placed between the walls of a slit pore of width $D$, where the walls are perpendicular to the $z$ axis.  Now we perform Monte Carlo simulations in the canonical ensemble ($NVT-MC$), by considering an $L_x\times L_x \times D$ square cuboid box subject to periodic boundary conditions along the transversal $x$ and $y$ directions. We take $L_x$ larger than the output from the $NPT-MC$ simulation in order to prevent interactions between the particles of a nanobridge and those located on its periodic images, but not so large that the nanobridge may evaporate. We usually consider runs of $10^6$ cycles to equilibrate the nanobridge, followed by other $10^6$ cycles to obtain the averages. In the $NVT-MC$ simulations a cycle is a set of $N$ attempts to simultaneously translate and rotate a particle chosen at random.

During the simulation we monitor different physical quantities, such as the mean potential energy and the global orientational ordering. For the latter, we quantify the global nematic ordering by first calculating the tensor order parameter:
\begin{equation}
\mathbf{Q}=\left\langle \frac{1}{N}\sum_{i=1}^N \frac{3{\bf u}_i\otimes{\bf
u}_i-\textrm{\bf I}}{2} \right\rangle \label{defq}
\end{equation}
and by then choosing the larger positive eigenvalue of the
tensor order parameter, $S$, as the representative measure of the
global nematic order. We note that, with this selection, the
director of the phase would correspond to the associated
eigenvector, $\mathbf{N}$.

As usual, we choose reduced units, where the energies and the lengths are
chosen in units of $\epsilon_0$ and $\sigma_0$, respectively. In particular
the reduced temperature $T^*$ and number density $\rho^*$ are defined as:
\begin{equation}
T^*=\frac{k_B T}{\epsilon_0}\qquad,\qquad \rho^*=\rho \sigma_0^3\ .
\label{reducedunits}
\end{equation}
Hereafter we will drop the asterisk, so quantities are given in reduced units.

\subsection{Evaluation of the density and order parameter profiles}

We also characterize the shape and nematic texture within the nanobridge
by calculating density and orientational profiles. The density profile, $\rho(x,y,z)$, is defined as:
\begin{equation}
\rho(x,y,z)\equiv \left\langle \sum_{i=1}^N \delta(x_i-x)\delta(y_i-y)\delta
(z_i-z) \right\rangle, \label{defrhoz}
\end{equation}
where $(x_i,y_i,z_i)$ are the Cartesian coordinates of the particle $i$, $\delta$ denotes Dirac's delta and $\langle \ldots \rangle$ stands for the ensemble average. The orientational order profile is
given by the local tensor order parameter
\begin{equation}
\mathbf{Q}(x,y,z)\equiv \frac{1}{\rho(x,y,z)}\left\langle \sum_{i=1}^N 
\frac{3{\bf u}_i\otimes{\bf u}_i-\textrm{\bf I}}{2} 
\delta(x_i-x)\delta(y_i-y)\delta(z_i-z)\right\rangle, \label{defqz}
\end{equation}
where $\mathbf{u}_i$ is the unit vector parallel to the symmetry axis of the particle $i$ and $\textrm{\bf I}$ is the identity matrix. The scalar nematic order parameter profile $S(x,y,z)$ and the local director field $\mathbf{n}(x,y,z)$ are obtained as the largest eigenvalue and the corresponding eigenvector of $\mathbf{Q}(x,y,z)$, respectively. It is also possible to get the biaxiality profile as the difference between the two smallest eigenvalues, but we see that this quantity does not provide further information, so we will skip it in our analysis.

For the numerical evaluation of the density profile and local tensor order parameter, we divide the box in a grid of cubic voxels of unit side length (in reduced units), so the density and local tensor order profiles are obtained by averaging in each voxel.. This voxel size is chosen as a compromise between coarse-graining and statistical precision, since smaller sizes are subject to larger statistical fluctuations, while larger sizes lead to a blurred picture of the density and order parameter profiles.

In some cases the nanobridges are axisymmetric around the $z$ axis. In this situation, the global nematic director $\mathbf{N}$ is also parallel to the $z$ axis. Under these circumstances, we can obtain the density profile and the local tensor order parameter by a similar procedure to that previously used for nanodrops in calamitic \cite{Rull2012} and oblate \cite{Rull2017} liquid crystals. This procedure generates density and orientational order parameter profiles with better resolution than in the procedure described above.
So, the density profile, $\rho(r,z)$, is defined as:
\begin{equation}
\rho(r,z)\equiv {\frac{1}{2\pi r}} \left\langle \sum_{i=1}^N \delta
(z_i-z)\delta(r_i-r) \right\rangle \label{defrhoz1}
\end{equation}
where $(r_i,z_i)$ are, respectively, the instantaneous radial and
vertical coordinates of particle $i$. Regarding the orientational order profile, 
the local tensor order parameter will present in different points of the 
$(r,z)$ shell eigenvectors which are related by a rotation around the 
$z$-axis. Then, the representation of the 
local tensor order parameter in the local cylindrical basis 
$(\mathbf{u}_r,\mathbf{u}_{\phi},\mathbf{u}_z)$ has as components
\begin{equation} 
Q_{\alpha\beta}(r,z)\equiv {\frac{1}{2\pi r\rho(r,z)}} 
\left\langle \sum_{i=1}^N 
\frac{3{u}_i^\alpha{u}_i^\beta-\delta_{\alpha,\beta}}{2} \delta
(z_i-z) \delta (r_i-r)\right\rangle \label{defrhoz2}
\end{equation}
where the Greek indexes represent the vector components in the cylindrical 
basis associated to the position of the particle. By diagonalization of the 
matrix defined by Eq.~(\ref{defrhoz2}) we get the local nematic order parameter profile 
$S(r,z)$ and the local nematic director field $\mathbf{n}(r,z)$ as 
its largest eigenvalue and associated eigenvector, respectively.
For its numerical implementation, now we divide the
nanobridge along the $z$ axis into cylinders
of height $\Delta z$. Every slice is further divided into cylindrical 
shells of average radius $r$ and width $\Delta r$.
We have used in this work $\Delta r=1/2$ and
$\Delta z=1/4$, again in reduced units. The local density is obtained by 
the quotient of the number of particles within each shell and its volume, and
the local tensor order parameter as the average over the particles on the shell 
of the matrix $(3{u}_i^\alpha{u}_i^\beta-\delta_{\alpha,\beta})/2$ (expressed in 
cylindrical coordinates). 

\section{Results}
\label{section3}

We report simulation results for a system of $N=32000$ GB oblate particles with $\kappa=0.5$ and $\kappa'=1$. With these parameters, nematic-vapour coexistence is observed in the narrow range of temperatures $T=0.4-0.5$. We choose $T=0.5$, in which the density of the nematic phase at coexistence is $\rho=1.86(2)$, with a nematic order parameter $S=0.50$ \cite{Rull2017,Rull2017_2}. In addition, the nematic-vapour interface shows homeotropic anchoring \cite{Rull2017_2}, as expected since $\kappa<\kappa'$. In absence of confining walls, near-spherical nanodroplets of nematic liquid crystals are observed under these conditions, with a disclination ring of topological charge $+1/2$ located inside the droplet on its equatorial plane \cite{Rull2017}. Now we are going to characterize the formation of nanobridges in slit pores. For this purpose, we have considered three different values of the adimensional wall-particle potential strength $a=0.25$, $a=0.6$ and $a=1$. 
\subsection{Nanodroplets on a single wall}
\begin{figure}
\centering
\subfloat{%
\resizebox*{6cm}{!}{\includegraphics{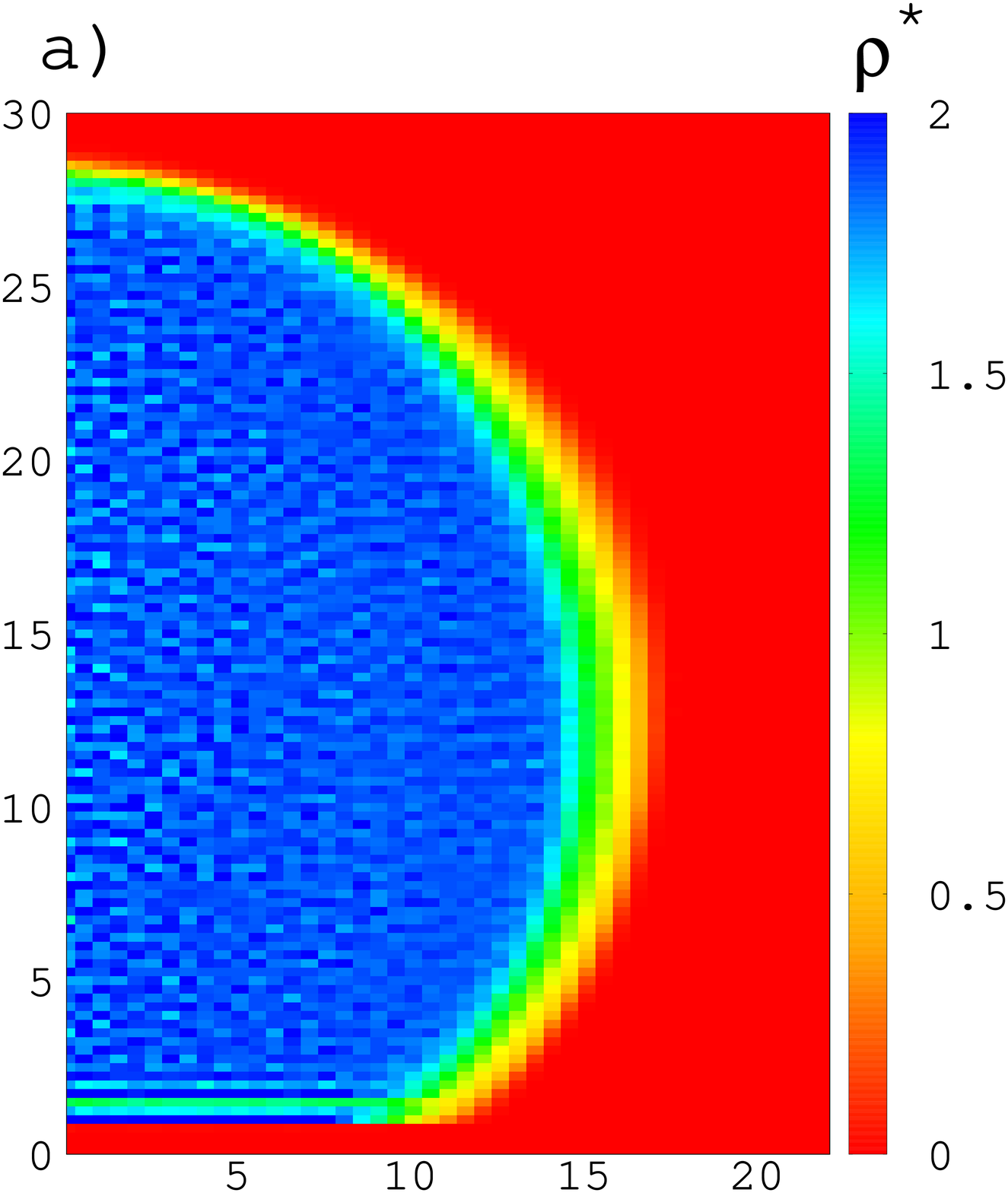}}}\hspace{5pt}
\subfloat{%
\resizebox*{6cm}{!}{\includegraphics{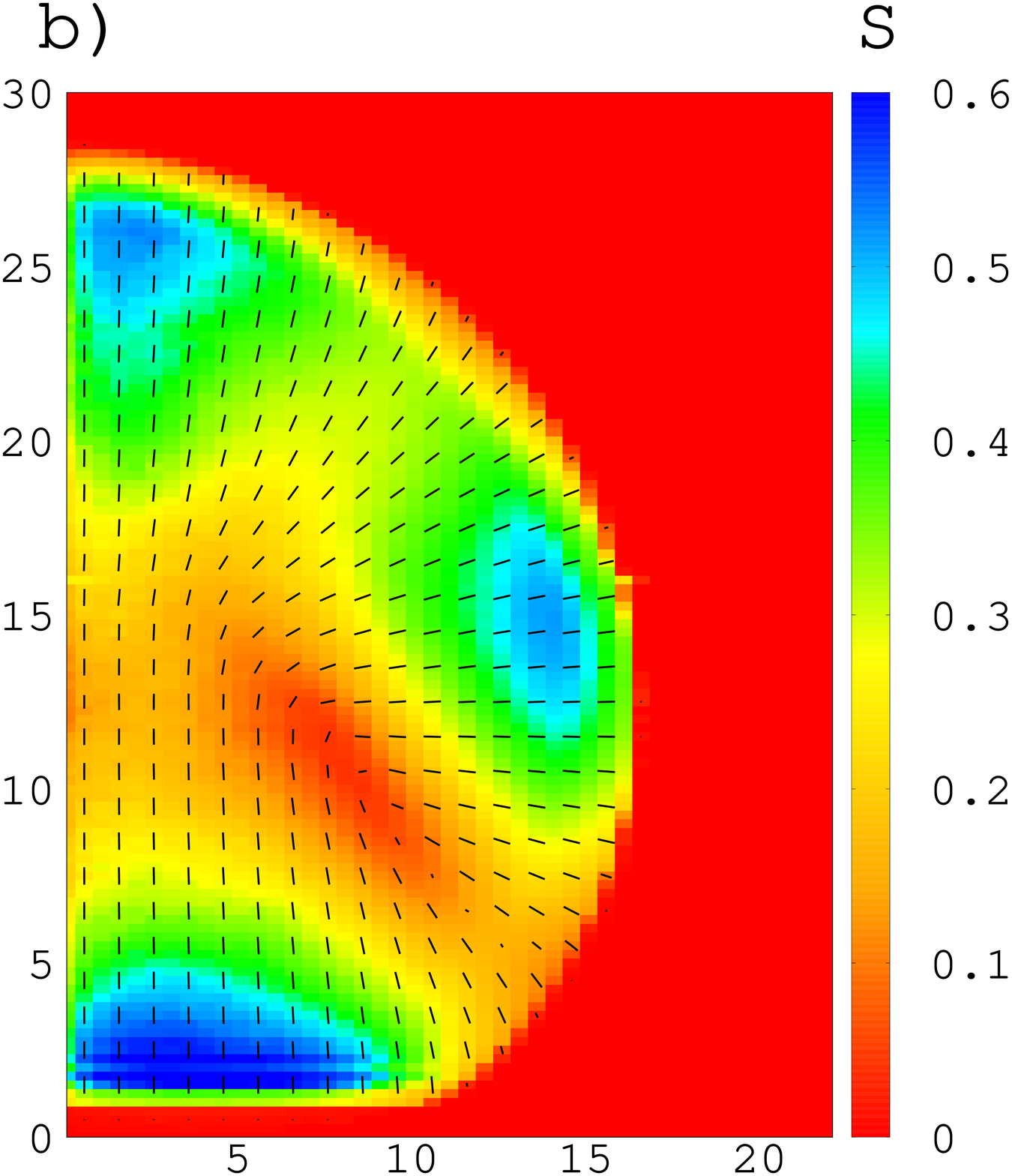}}}\vspace{-1.7cm}\\
\subfloat{%
\resizebox*{6cm}{!}{\includegraphics{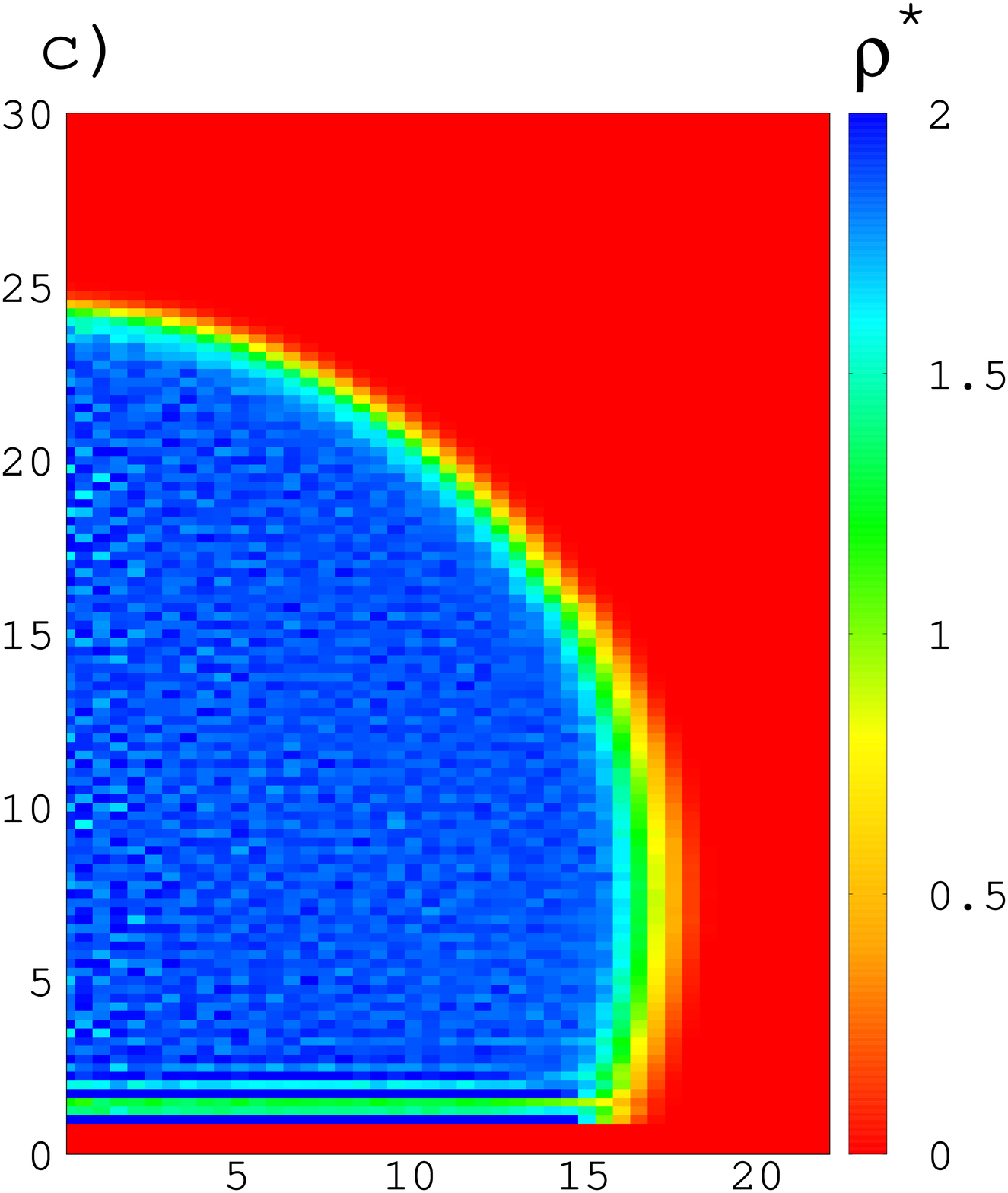}}}\hspace{5pt}
\subfloat{%
\resizebox*{6cm}{!}{\includegraphics{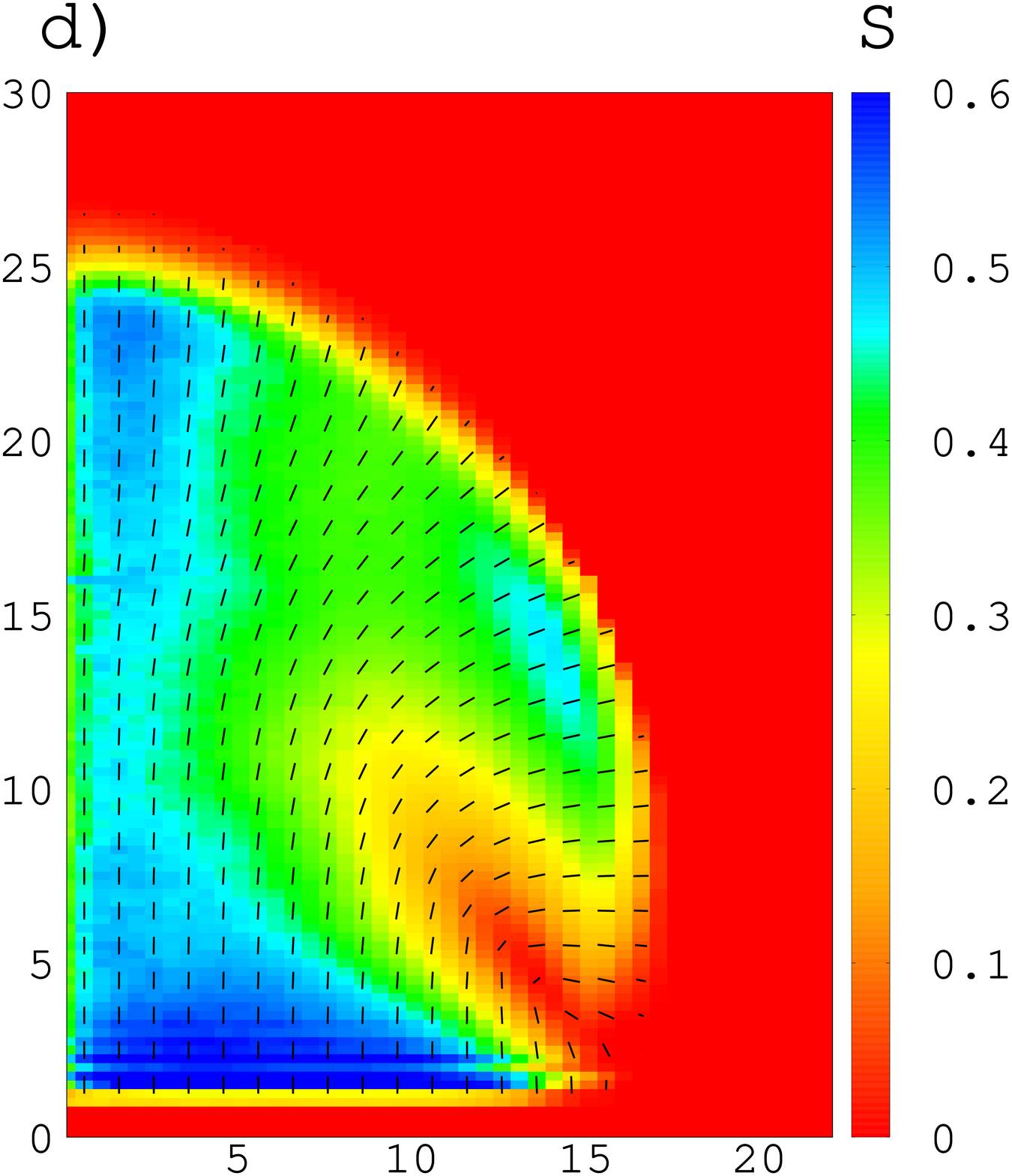}}}\vspace{-1.7cm}\\
\subfloat{%
\resizebox*{6cm}{!}{\includegraphics{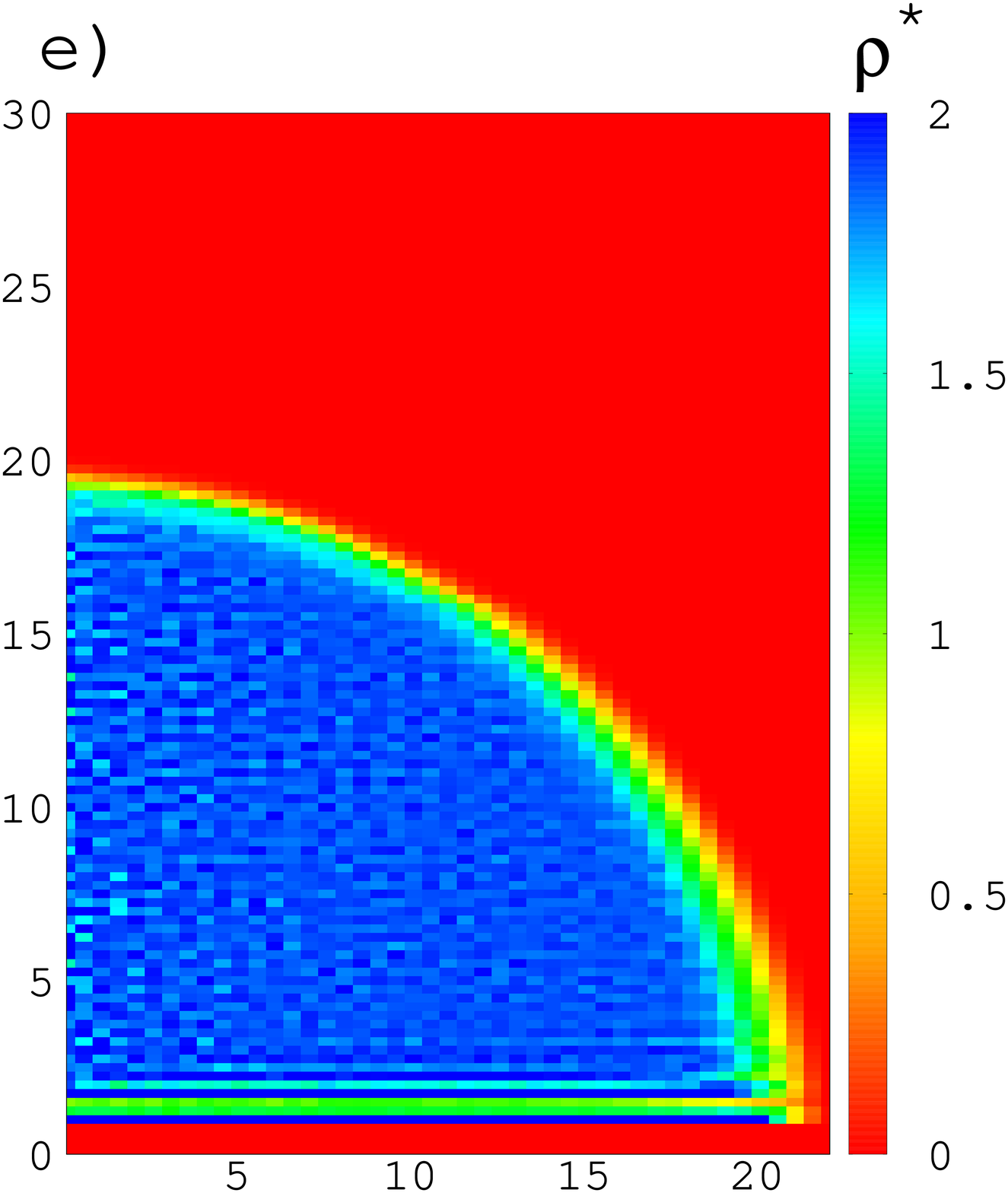}}}\hspace{5pt}
\subfloat{%
\resizebox*{6cm}{!}{\includegraphics{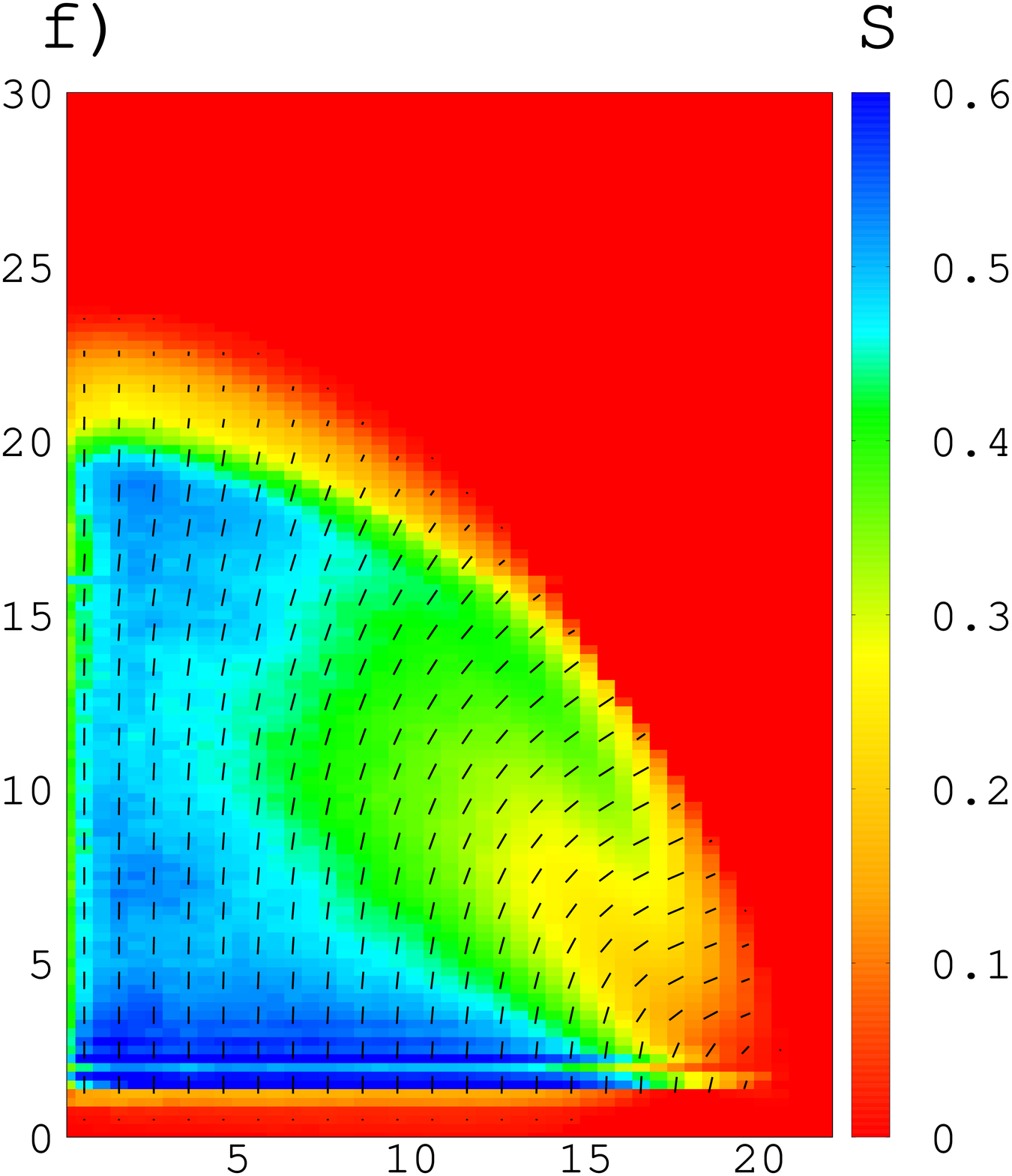}}}
\caption{Plots of the density profiles (left column) and orientational order profiles (right column) of a nanodrop of nematic liquid crystal on top of a wall with $a=0.25$ (top row), $a=0.6$ (middle row) and $a=1.0$ (bottom row). In each figure the abscissa axis corresponds to the radial coordinate $r$ and the ordinate axis to the vertical coordinate $z$. For the density $\rho$ and scalar order parameter $S$ profiles a colour mapping is considered. In addition, on top of the $S$ profile representation, segments represent the local nematic director orientation.} \label{fig1}
\end{figure}
As a previous analysis, we studied the wetting properties of a single wall. For this purpose, we modified the procedure to generate nanobridges to study the formation of a sessile nanodroplet on a wall. In order to do that, we placed the outcome of the bulk $NPT-MC$ simulation on one of the walls which limit the slit pore, with a width much larger than side length of the $NPT-MC$ simulation on the $z$ direction. For all values of the potential strength $a$, nanodroplets are shaped as near-spherical caps on the wall. This observation is consistent with the almost spherical morphology of the nematic nandroplets \cite{Rull2017}. In addition, the order parameter profile is axisymmetric around the $z$ axis and there is no quirality, i.e. the azimuthal component of the local director vanishes everywhere. Thus, we characterize the density and orientational order parameter profiles by Eqs.~(\ref{defrhoz1}) and (\ref{defrhoz2}). Our simulation results are summarized in Fig.~\ref{fig1}.  The density is quite uniform and takes approximately the bulk value inside the droplet. Deviations are only observed in the nematic-vapour interface and close to the wall, which is located at $z=0$. The density variation through the nematic-vapour interface is smooth and consistent with the results reported in Ref.~\cite{Rull2017_2}. On the other hand, close to the wall there is an empty layer of about one unity and, after that, it is observed some density layering which decays to the bulk value. This observation is consistent with the fact that the single wall-particle potential has a highly repulsive region for $z<\sigma_w/\sigma_0 \sim 1$, while the layering is the typical for dense fluids close to hard walls due to depletion effects. Regarding the shape of the droplet, it fits quite nicely to a near-spherical cap, characterized by a single (apparent) contact angle $\theta$ between the wall portion in contact with the nematic and the nematic-vapour interface at the nanodroplet contact point. This contrasts with previous studies of wetting sessile nematic droplets of calamitic nematogens on flat walls, which are anisotropic \cite{Vanzo20161610}. We estimate that $\theta\approx 144^\circ$ for $a=0.25$, $\theta\approx 114^\circ$ for $a=0.6$ and $\theta\approx 85^\circ$ for $a=1.0$, meaning that the wall is nematic-phobic for $a=0.25$ and $a=0.6$, or slightly nematic-philic for $a=1.0$. The decrease of the contact angle with the potential strength $a$ is expected, since in wetting by liquids usually the more attractive the wall-particle potential is, the more hygrophilic the substrate is. However, it is important to note that these apparent contact angles may differ from the true thermodynamic contact angle, as large finite-size deviations are observed even in simple fluids \cite{Santiso2013}, which may be enhanced by effect of the elastic distortions within the droplet.

Regarding the orientational order parameter, we see that the nematic phase preferentially anchors homeotropically on the nematic-isotropic interface, where the scalar nematic order parameter $S$ decreases smoothly to zero through the interface, in agreement with previous studies by the authors \cite{Rull2017_2}. The wall also shows homeotropic anchoring but now the local scalar nematic order parameter is enhanced in the layers of higher density. Thus, there are elastic distortions inside the nanodroplet to adapt to these anchoring conditions on the boundaries of the nanodroplet. For $a=0.25$, a $+1/2-$disclination ring is observed at $z\sim 10$, similar to the one observed in the free nanodroplet \cite{Rull2017}. However its core, defined as the region inside the nanodroplet with $S<0.1$, is highly distorted and the scalar nematic order parameter is in general reduced with respect to the bulk value, although biaxiality is restricted to the disclination core region (not shown). This indicates that the disclination ring wanders inside the droplet, as it also happened in the free nanodroplet. As $a$ is increased, the equilibrium position of the defect core moves towards the contact line of the sessile nanodroplet to the wall. For $a=0.6$ the disclination core is still quite elongated, but now the central part of the nanodroplet shows a more uniform value of $S$, approximately equal to the bulk value, where the nematic director field is almost uniformly vertical. For $a=1.0$, the disclination core has almost merged with the interfacial region but still the order is slightly reduced with respect to the bulk value in the neighbourhood of the nanodroplet-wall contact line. The central region with near-uniform, vertical nematic director, on the other hand, is enlarged. 
The attraction of the disclination core to the contact line can be understood if we consider that, for contact angles $\theta> 90^\circ$, the nematic director field close to it resembles the orientational field of a negative topological charge disclination ring, in a similar way as it happens for a nematic confined in a wedge \cite{Barbero1980,JMRE2010}. Thus, the bulk $+1/2$ disclination ring will be attracted to the contact line in a similar way as two disclination lines of opposite topological charge. However, if we impose that the anchoring is homeotropic both at the wall and the nematic-vapour interface, we find that its effective topological charge is $-(180^\circ/\theta-1)$ \cite{JMRE2010}. Thus, the effective attraction between the bulk disclination ring and the contact line increases as $\theta$ decreases, as it is proportional to the product of their topological charges. It is worth mentioning that, for $\theta$ is smaller that $90^\circ$, another texture with positive effective topological charge $+1$ associated to the contact line becomes more favourable energetically. However, around $\theta=90^\circ$ both are energetically almost equivalent. 

\subsection{Nanobridges in the slit pore}
Now we turn to the case of the formation of nanobridges in slit pores. As the number of particles $N$ is fixed in our simulations, its aspect ratio $\Gamma=d/D$ (where $d$ is the nanobridge diameter at its middle plane and $D$ is the pore width) decreases as $D$ increases. Note that, roughly, the nanobridge has a cylindrical shape, so $d\sim \sqrt{N/D}$. Thus, if $D$ is very large, the bridge will breakdown as $d$ becomes smaller of one molecular diameter. This value also depends on the wettability of the walls. So, for example, no nanobridges are observed for $a=0.25$ for $D>40$ since the nematic detaches from the wall to form a free droplet inside the slit pore. On the other hand, no bridges are observed for $a=2.0$ for $D>30$ since two independent sessile droplets are formed on each wall. Therefore, in the range of values of $a$ we considered, we will focus on the cases $D=20$, 30 and 40, which, as we will see, correspond to $\Gamma >1$, $\Gamma \approx 1$ and $\Gamma <1$ cases, respectively. For this cases, the lateral size of the simulation box is taken to be $L_x=44$, which is enough to avoid interactions between any nanobridge and its periodic replicas.
\subsubsection{D=20}
\begin{figure}
\centering
\includegraphics{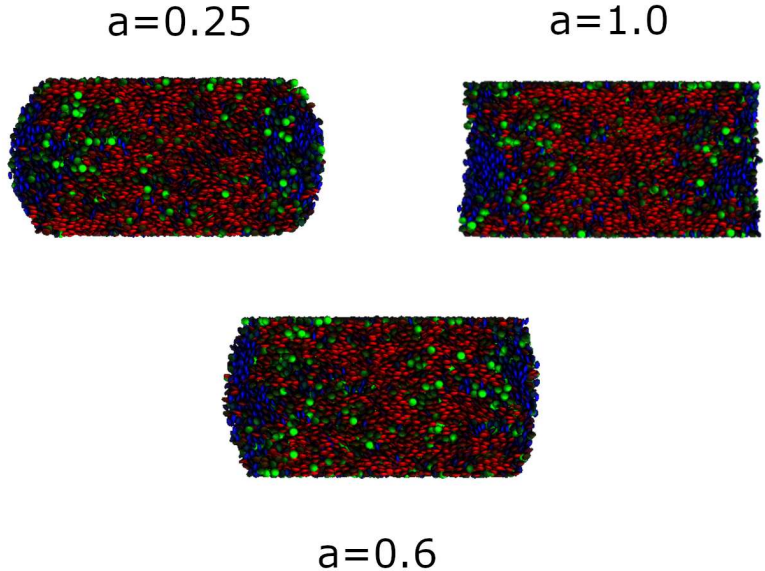} 
\caption{Snapshots of the cross section of a nanobridge in a slit pore of two horizontal walls separated by a distance $D=20$ and values of the wall-particle potential strength $a=0.25$, $a=0.6$ and $a=1.0$. The colour code associated to the GB particle orientations is the following: red if the main symmetry axes of the particles are aligned with the $z$ axis, blue if they are aligned with the $y$ (i.e. horizontal, in-plane) axis and green if they are aligned with the $x$ (i.e. horizontal, out-of-plane) axis.} \label{fig2}
\end{figure}

\begin{figure}
\centering
\subfloat{%
\resizebox*{7cm}{!}{\includegraphics{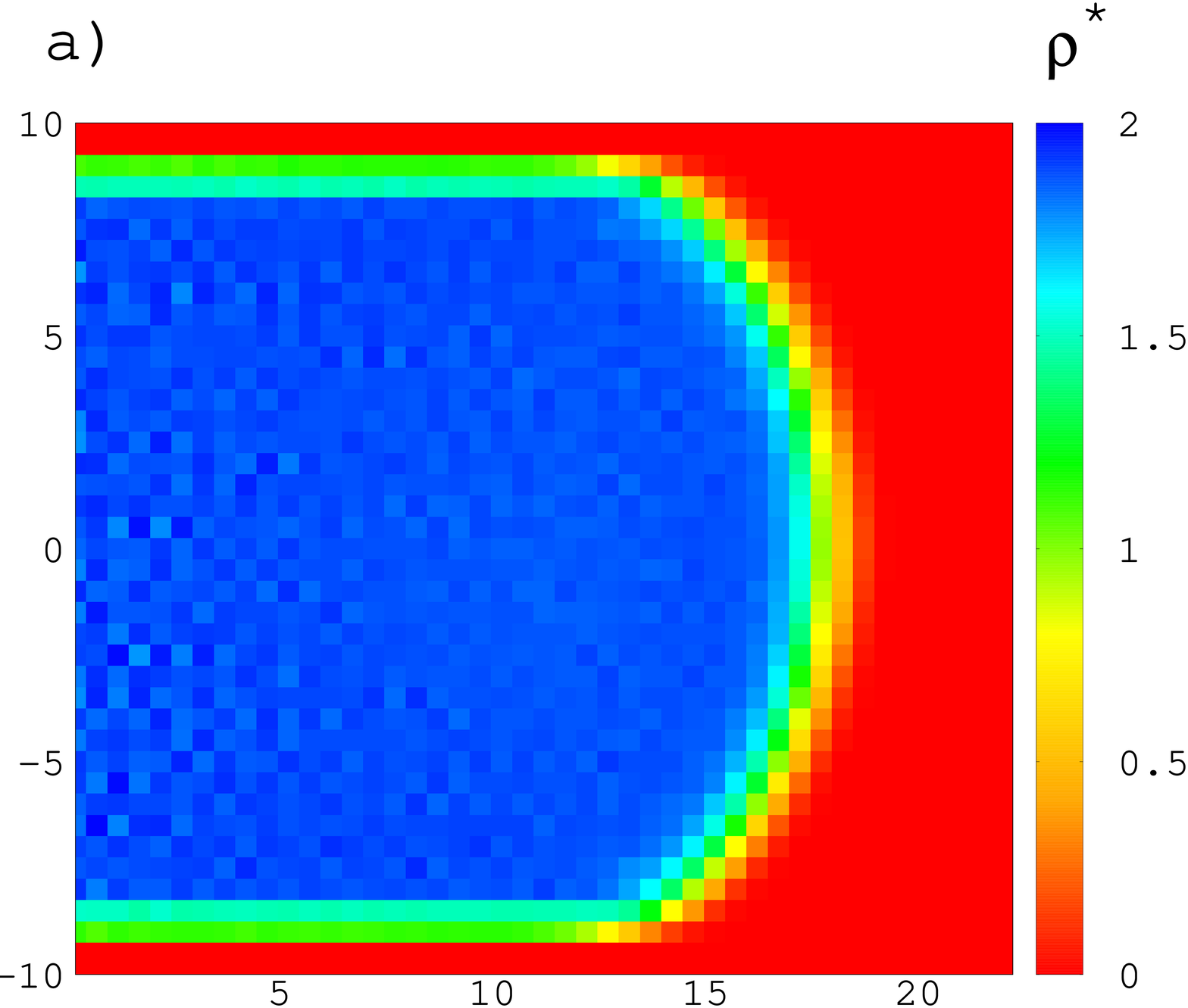}}}\hspace{5pt}
\subfloat{%
\resizebox*{7cm}{!}{\includegraphics{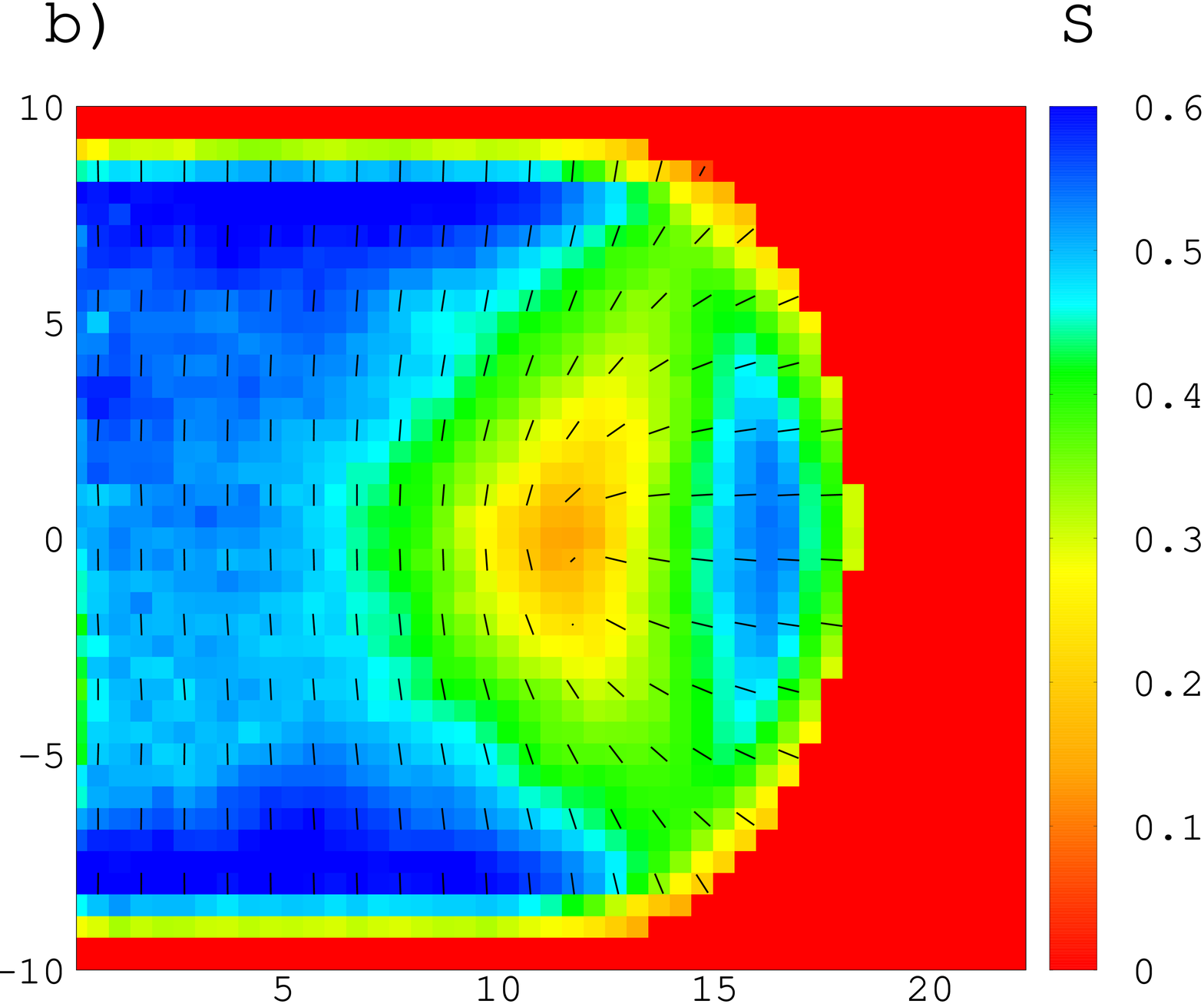}}}\vspace{-1.5cm}\\
\subfloat{%
\resizebox*{7cm}{!}{\includegraphics{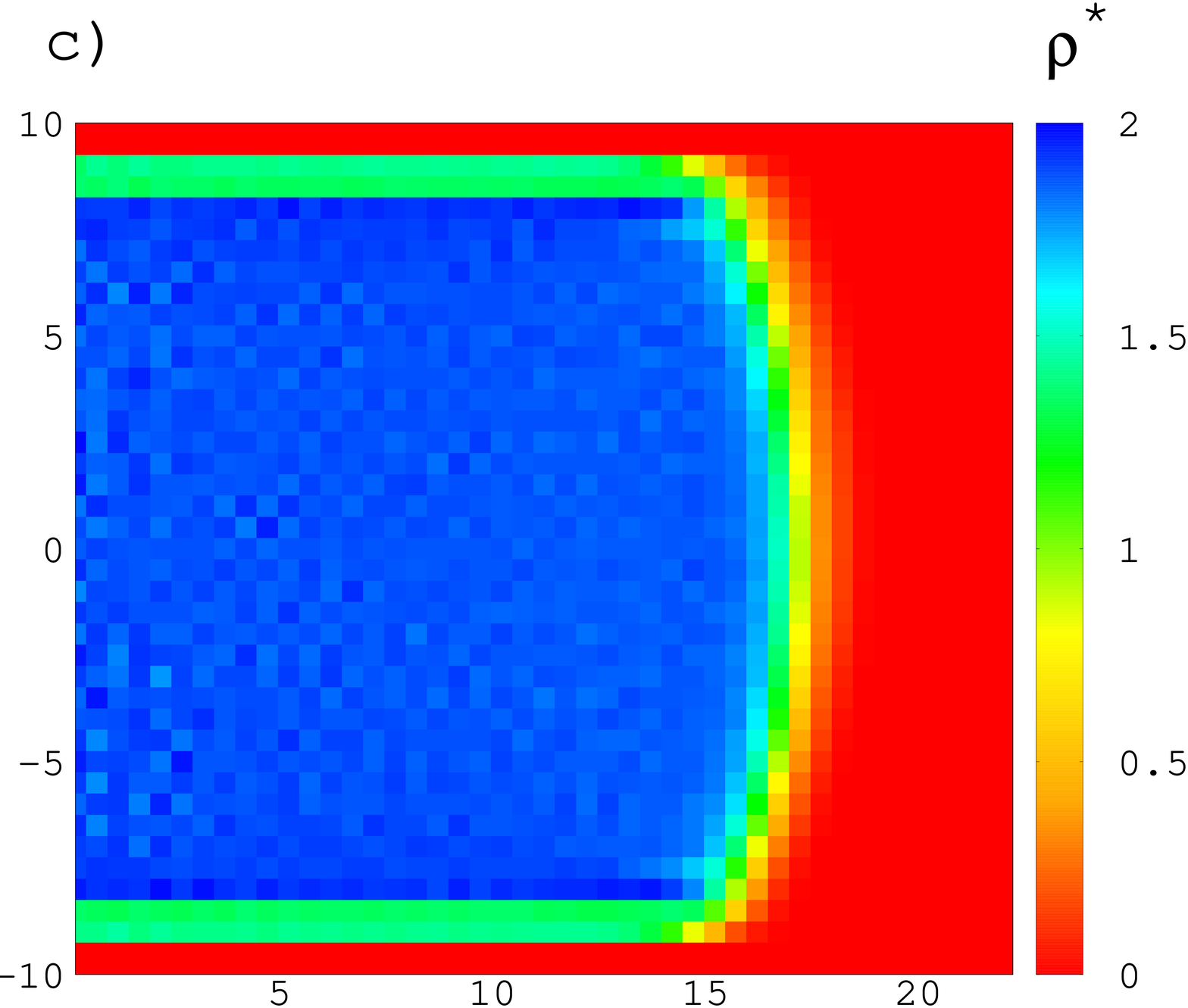}}}\hspace{5pt}
\subfloat{%
\resizebox*{7cm}{!}{\includegraphics{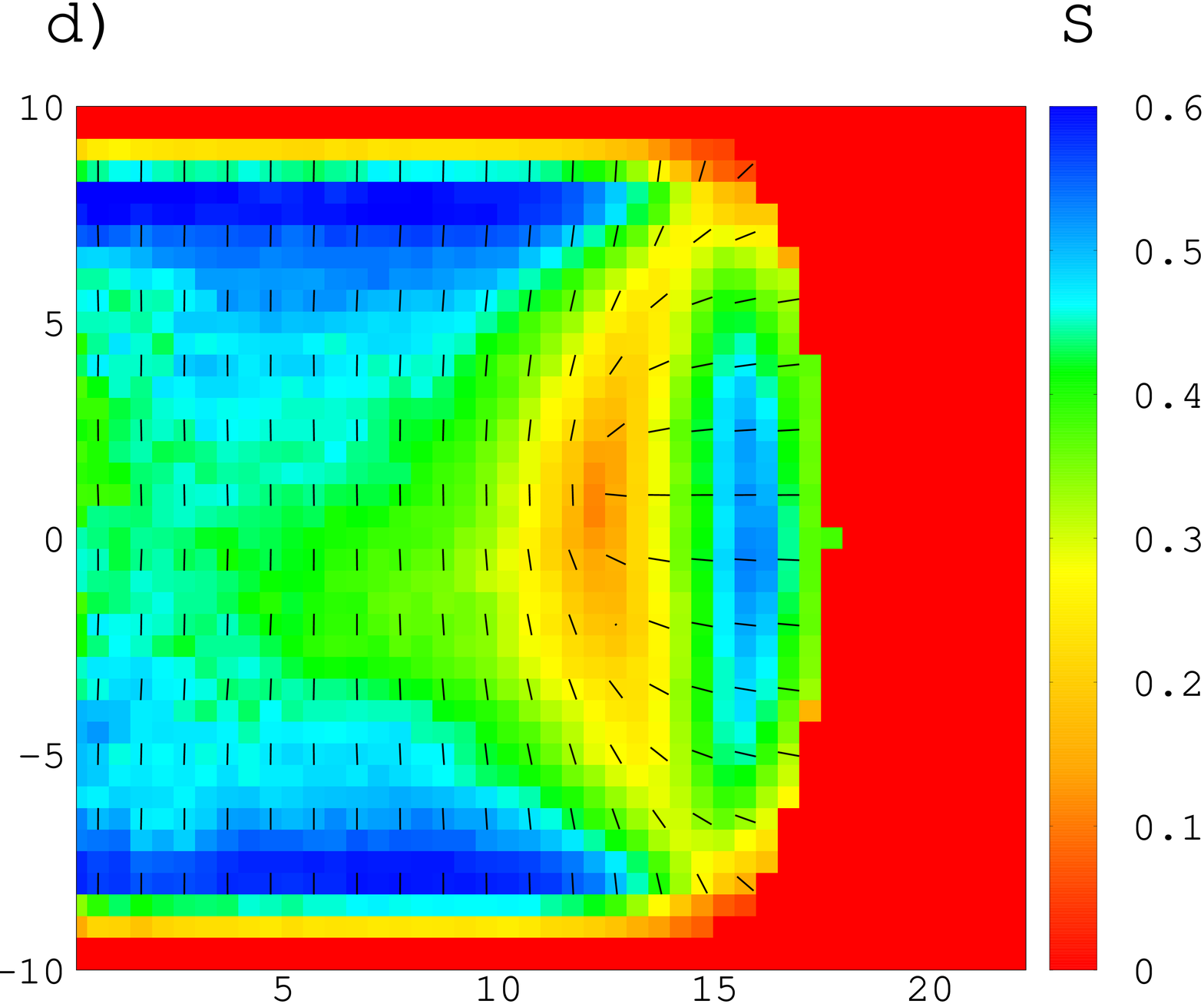}}}\vspace{-1.5cm}\\
\subfloat{%
\resizebox*{7cm}{!}{\includegraphics{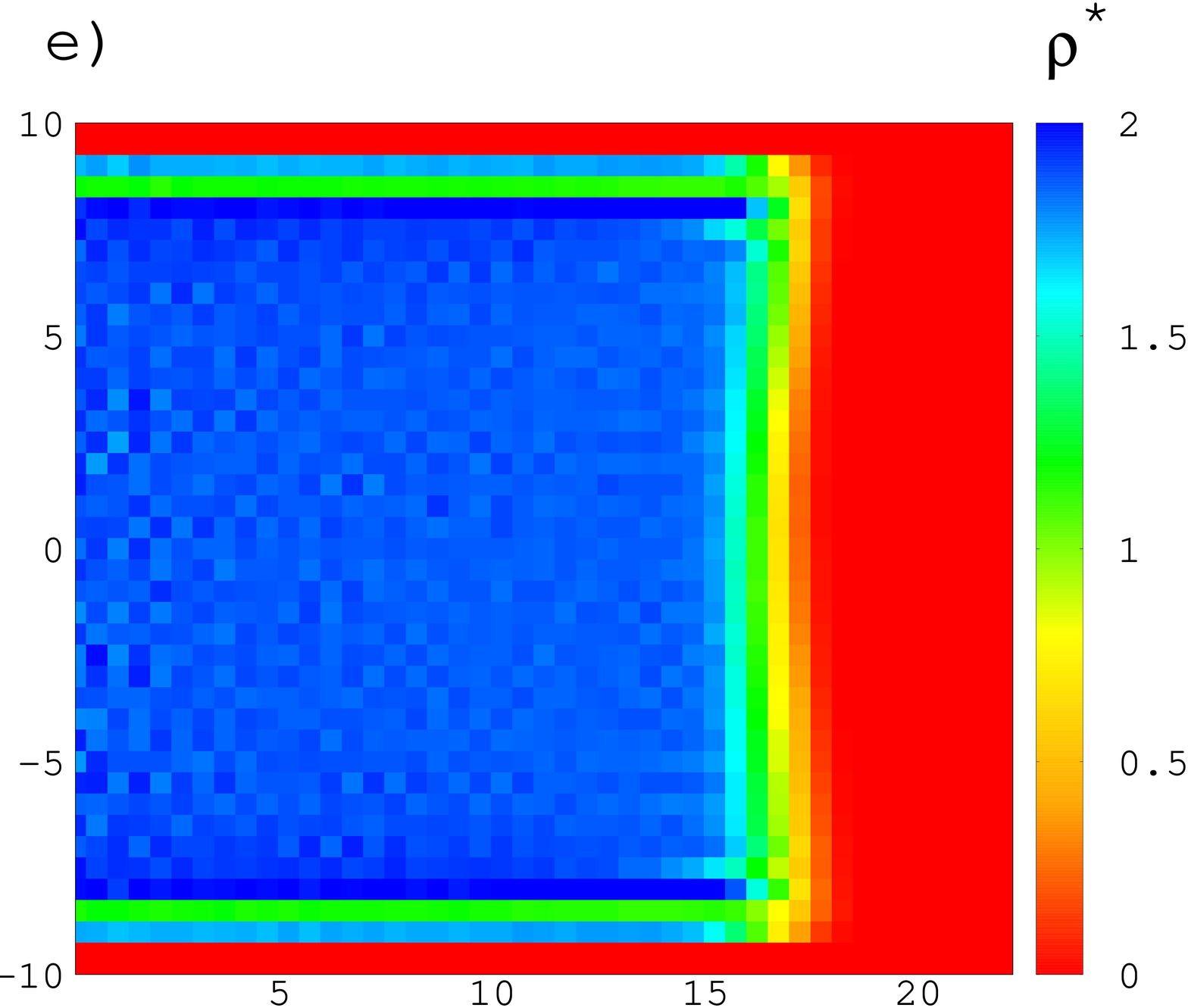}}}\hspace{5pt}
\subfloat{%
\resizebox*{7cm}{!}{\includegraphics{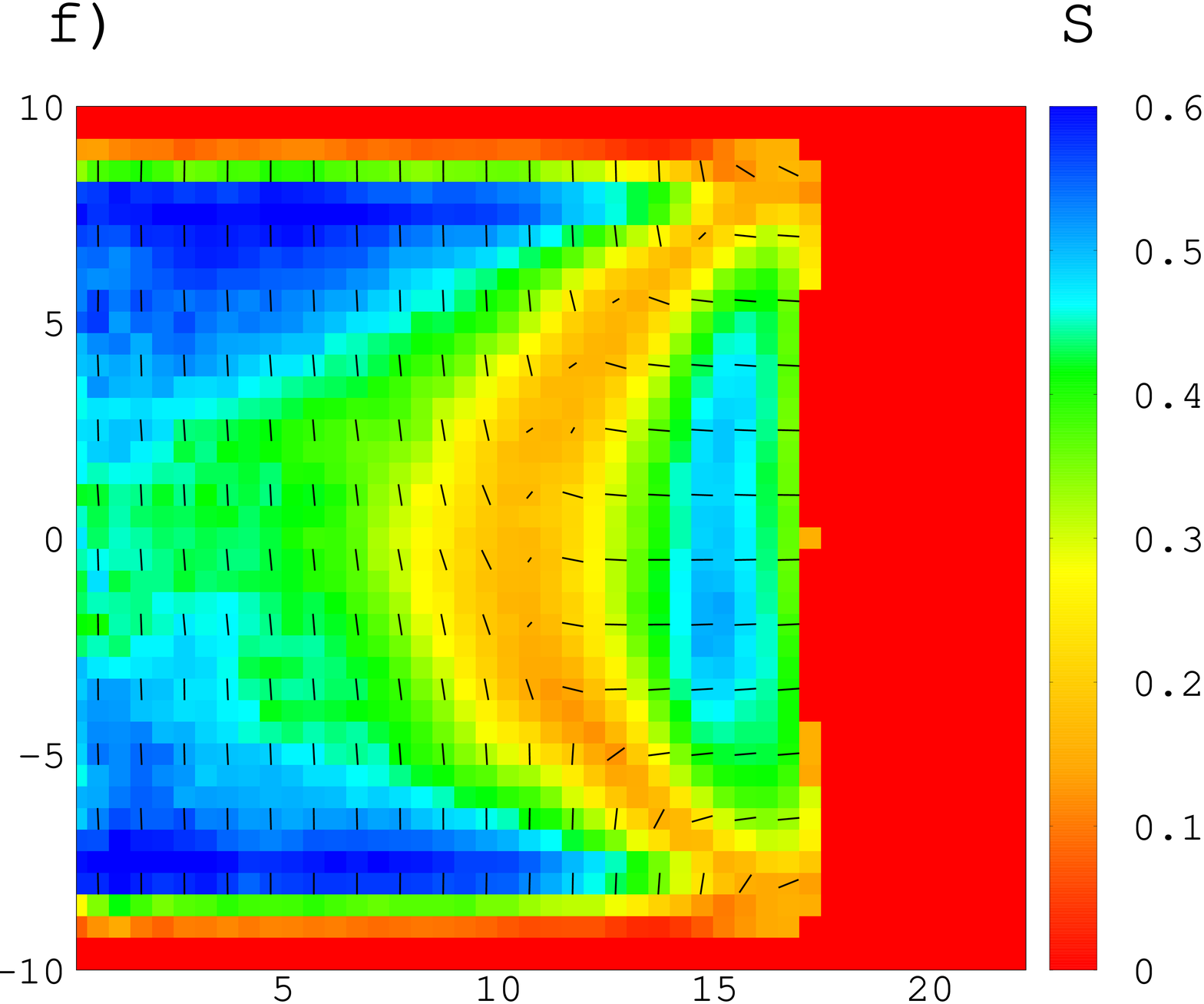}}}\vspace{0cm}
\caption{Plots of the density profiles (left column) and orientational order profiles (right column) of a nanodrop of nematic liquid crystal on top of a wall with $a=0.25$ (top row), $a=0.6$ (middle row) and $a=1.0$ (bottom row), where we used the same representation as in Fig.~\ref{fig1}.} \label{fig3}
\end{figure}

For $D=20$ the aspect ratio of the nanobridges is $\Gamma<1$ for all the values of $a$. However, their shape depends on the wettability of the walls: the nanobridges are barrel-shaped for $a=0.25$ and $a=0.6$, and hourglass-shaped for $a=1.0$. Inspection of simulation snapshots (see Fig.~\ref{fig2}) indicate that particles are mainly oriented perpendicular to the walls throrought the nanobridge, except in a toroidal region close to the nematic-vapour interface, where particles orient perpendicular to the interface. Thus, the orientational order is axisymmetric around the vertical axis, with no hint of quirality. These observations are confirmed by the evaluation of the density and orientational order parameter profiles, shown in Fig.~\ref{fig3}, which, as in the case of the sessile nanodroplets, we evaluate by using Eqs.~(\ref{defrhoz1}) and (\ref{defrhoz2}). The density inside the nanobridges is almost homogeneous and takes the bulk value except close to the walls and nematic-isotropic interface, where the density profiles behave as it was described above in the case of the sessile nanodroplets. This is a common feature for all the values of $D$ we consider. Regarding the orientational order parameter profile, for $a=0.25$ we see that the scalar nematic order parameter profile takes uniformly a value approximately equal to the bulk value in the central part of the nanobridge, with an almost vertical nematic director field. On the equatorial plane and close to the nematic-vapour interface, we observe a $+1/2$ disclination ring which allows the director field to anchor homeotropically on the interface. The values of $S$ are slightly reduced with respect to the bulk value in a relatively extensive region around the disclination core which connects to the nematic-vapour interface. As the value of $a$ increases, the disclination core stretches towards the contact lines of the nanobridge with the walls, so that for $a=1.0$ the $S$ profile shows almost a continuous line joining the contact lines in which the nematic director changes abruptly from a vertical to an horizontal orientation as the nematic-vapour interface is approached. This observation can be explained by the same attraction that the disclination ring feels towards the wall-nanobridge contact lines considered previously in the case of sessile nanodroplets. Therefore, the disclination ring inside the nanobridge may wander towards the contact lines, leading to this extension of the region where the scalar order parameter is decreased with respect to the bulk value. 
\subsubsection{D=30}
\begin{figure}
\centering
\includegraphics{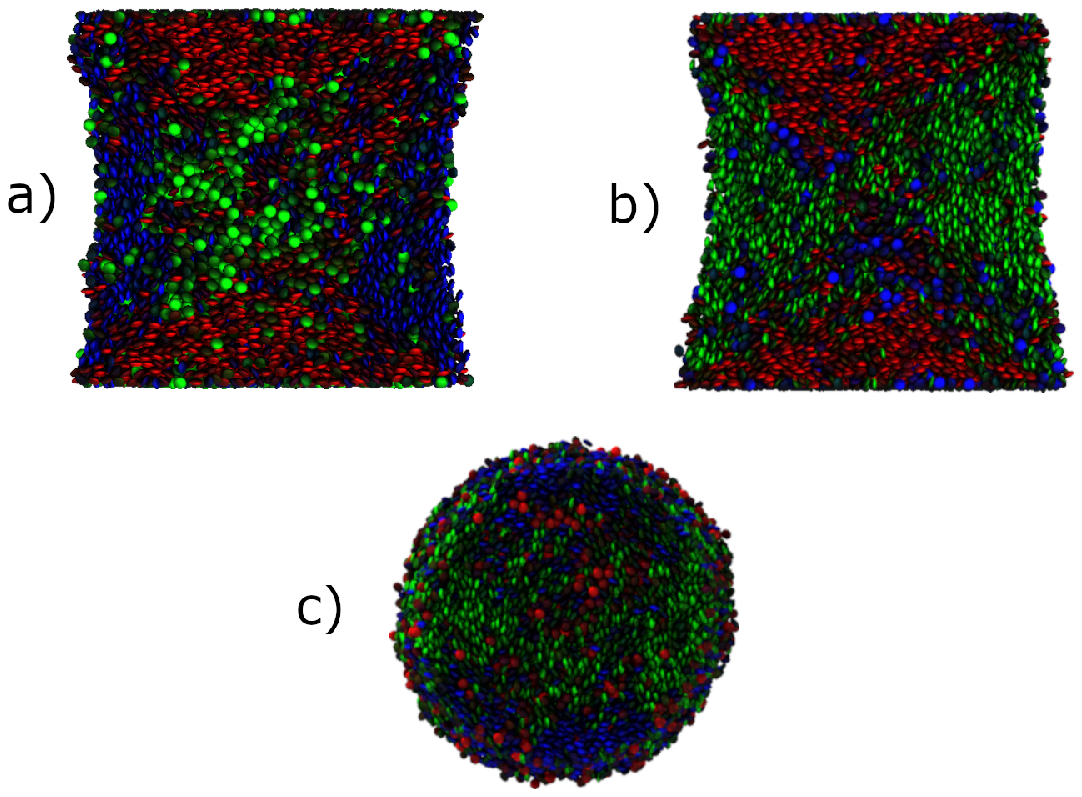}
\caption{Snapshots of the sections of a nanobridge in a slit pore of two horizontal walls separated by a distance $D=30$ and wall-particle potential strength $a=1.0$: a) $xz$ plane, b) $yz$ plane, and c) $xy$ plane. Colour code associated to the GB particle orientations is the same as in Fig.~\ref{fig2}.} \label{fig4}
\end{figure}
\begin{figure}
\centering
\subfloat{%
\resizebox*{6cm}{!}{\includegraphics{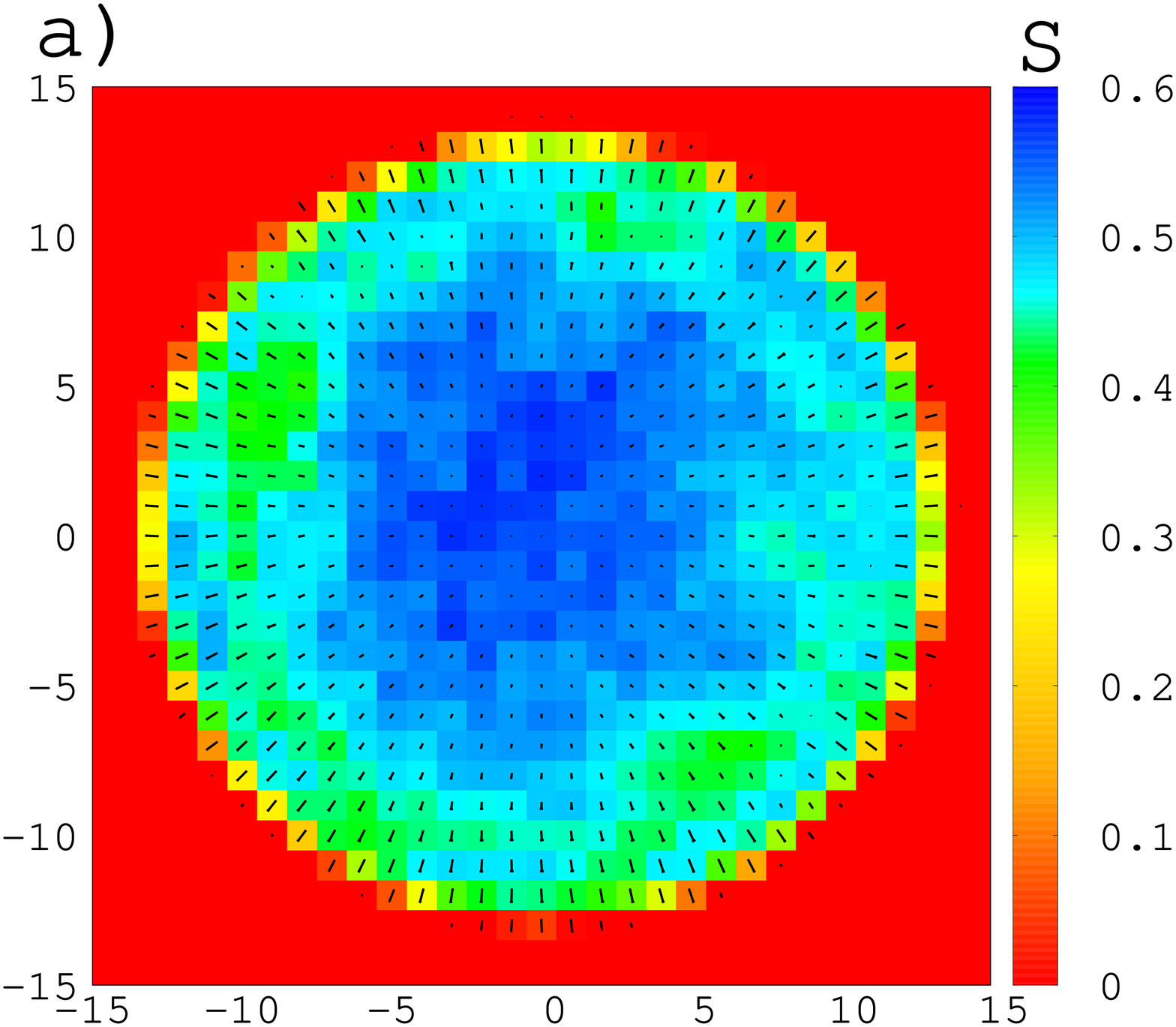}}}\hspace{5pt}
\subfloat{%
\resizebox*{6cm}{!}{\includegraphics{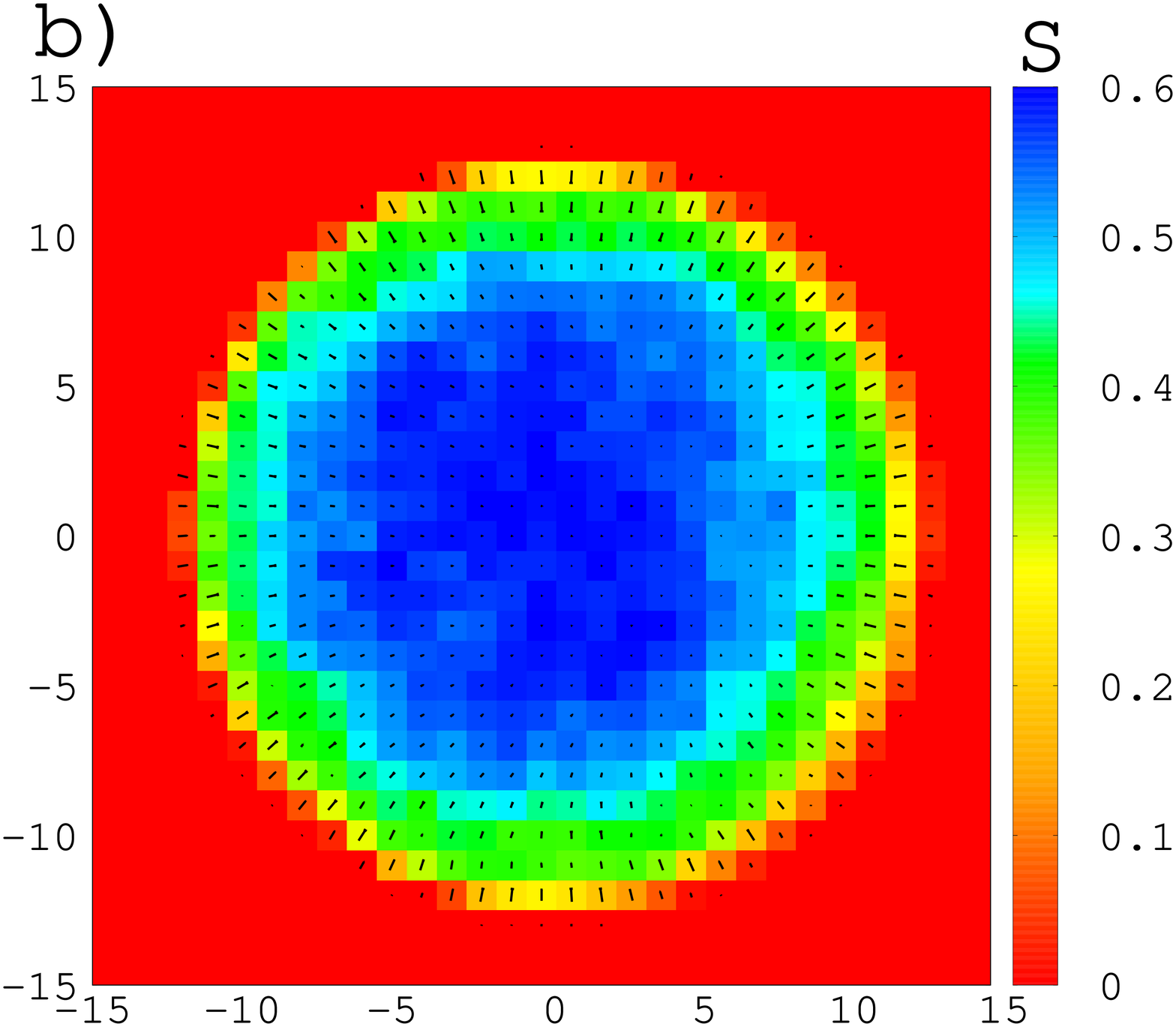}}}\vspace{-3.5cm}\\
\subfloat{%
\resizebox*{6cm}{!}{\includegraphics{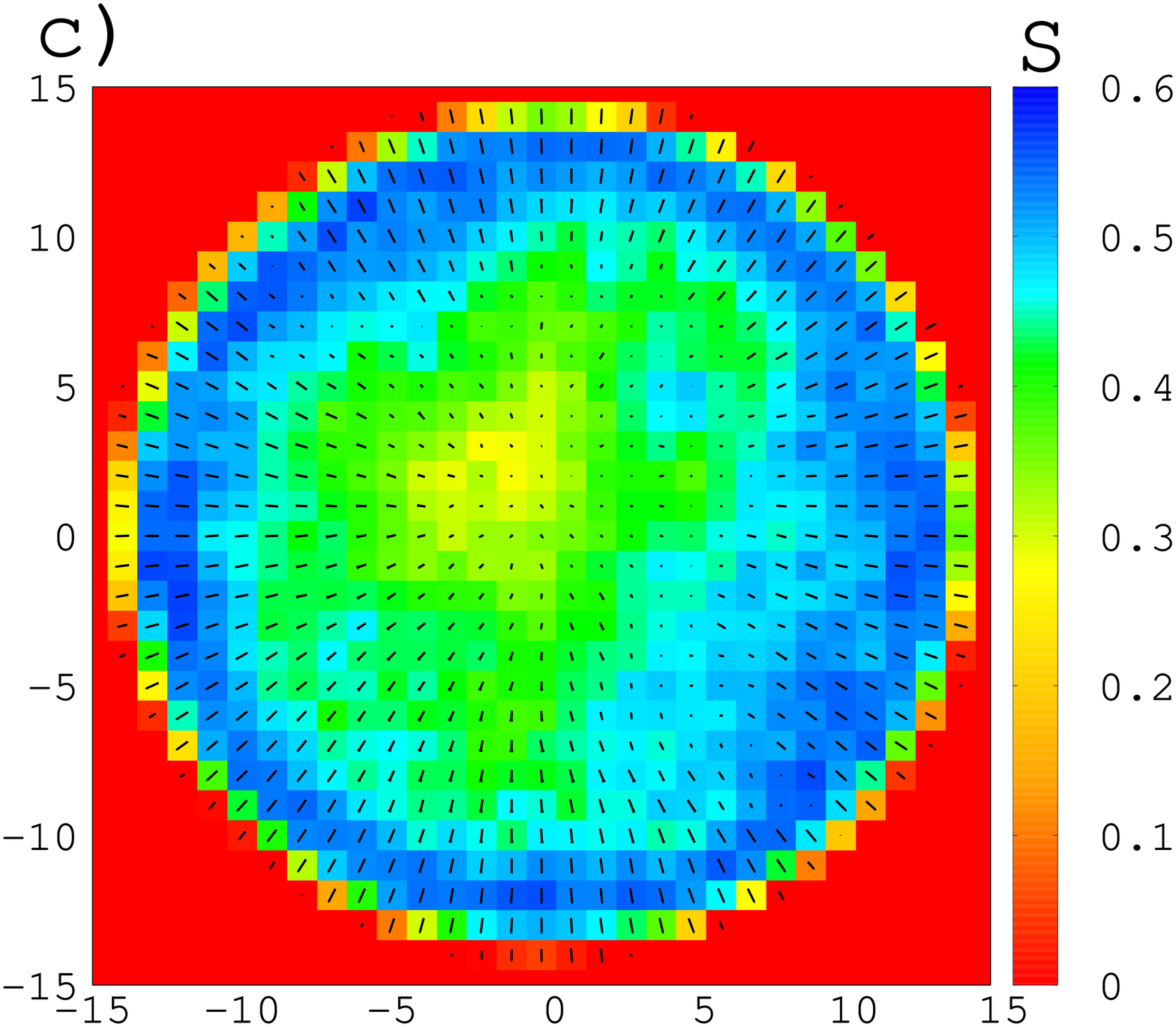}}}\hspace{5pt}
\subfloat{%
\resizebox*{6cm}{!}{\includegraphics{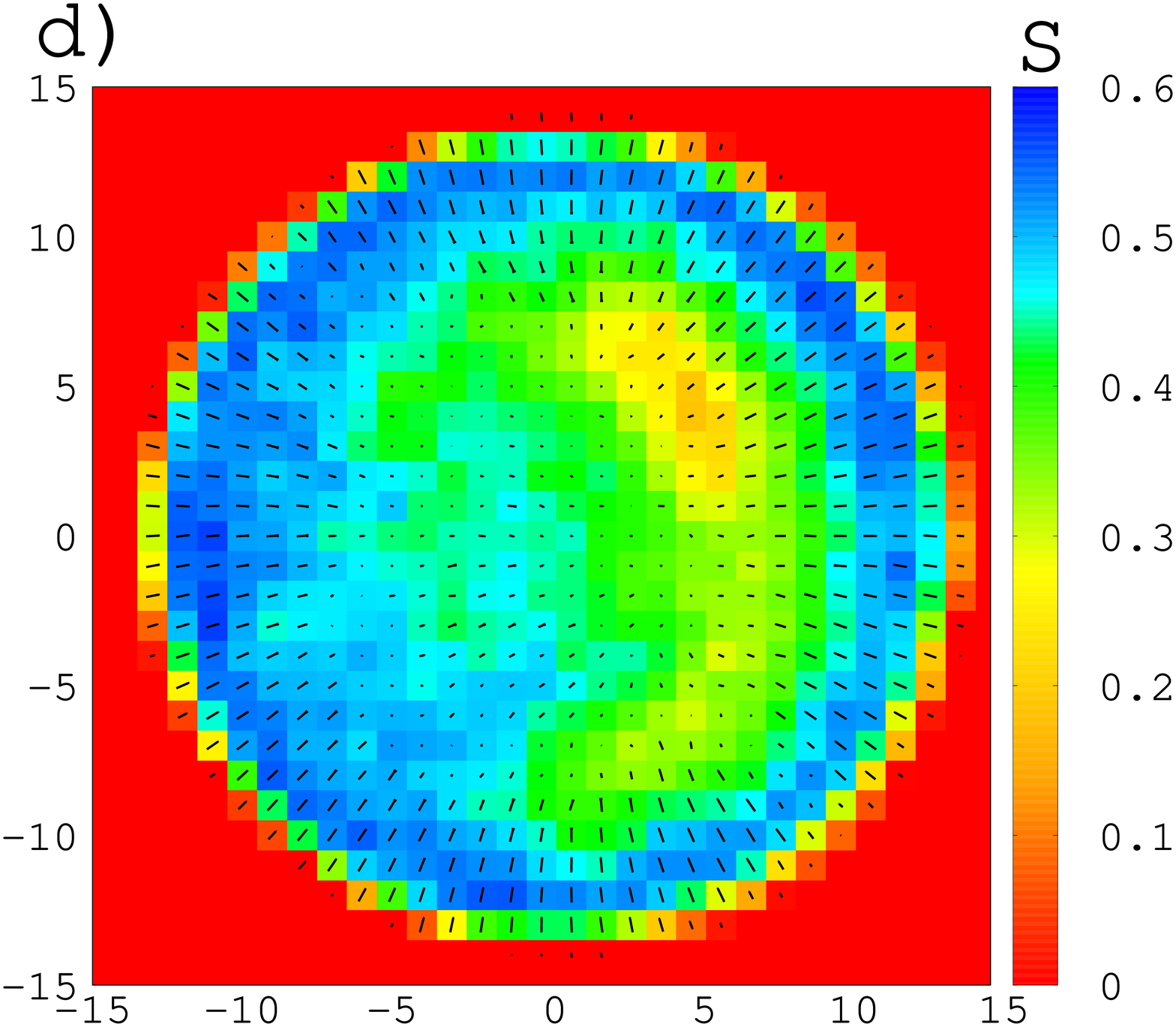}}}\vspace{-3.5cm}\\
\subfloat{%
\resizebox*{6cm}{!}{\includegraphics{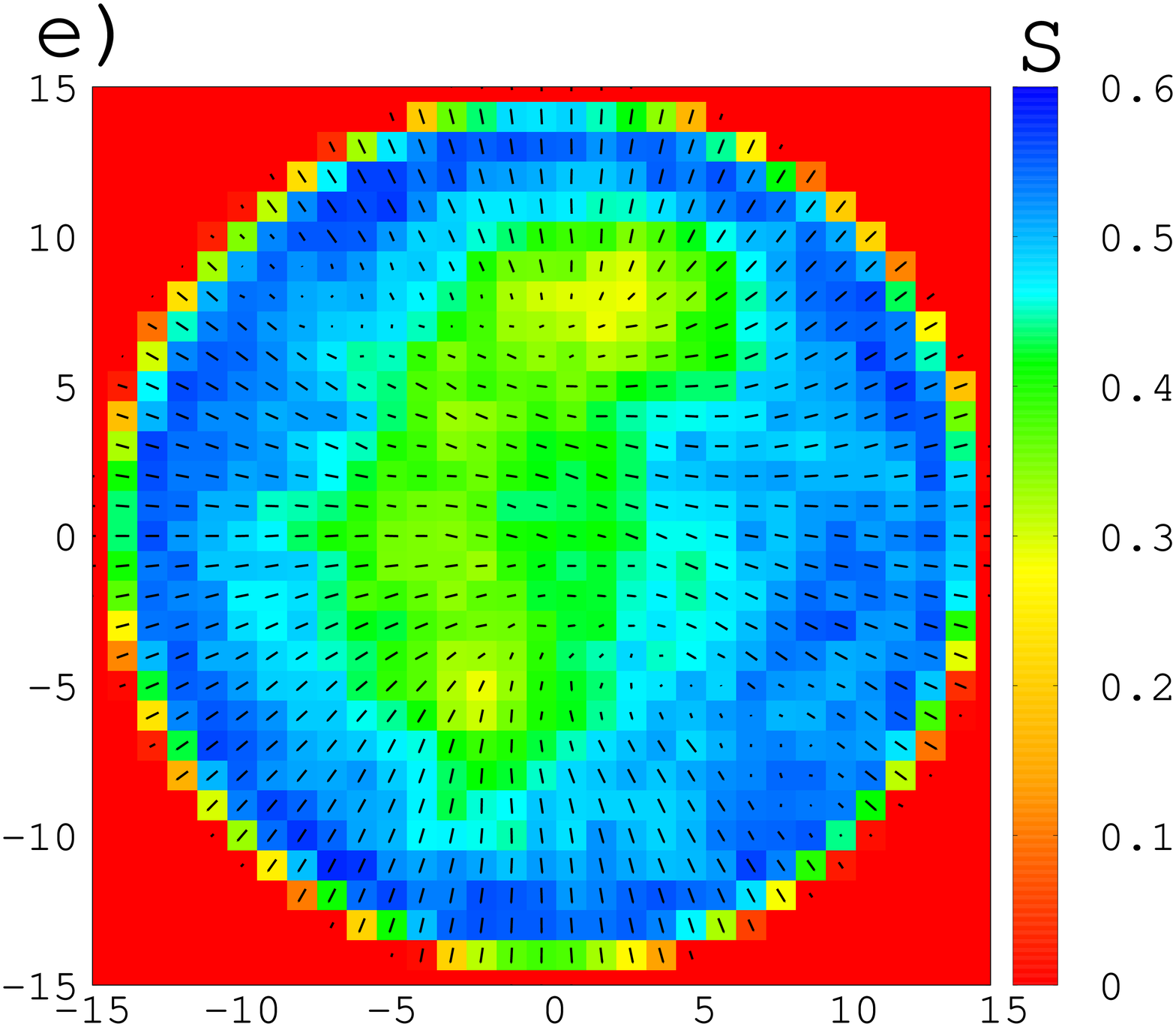}}}\hspace{5pt}
\subfloat{%
\resizebox*{6cm}{!}{\includegraphics{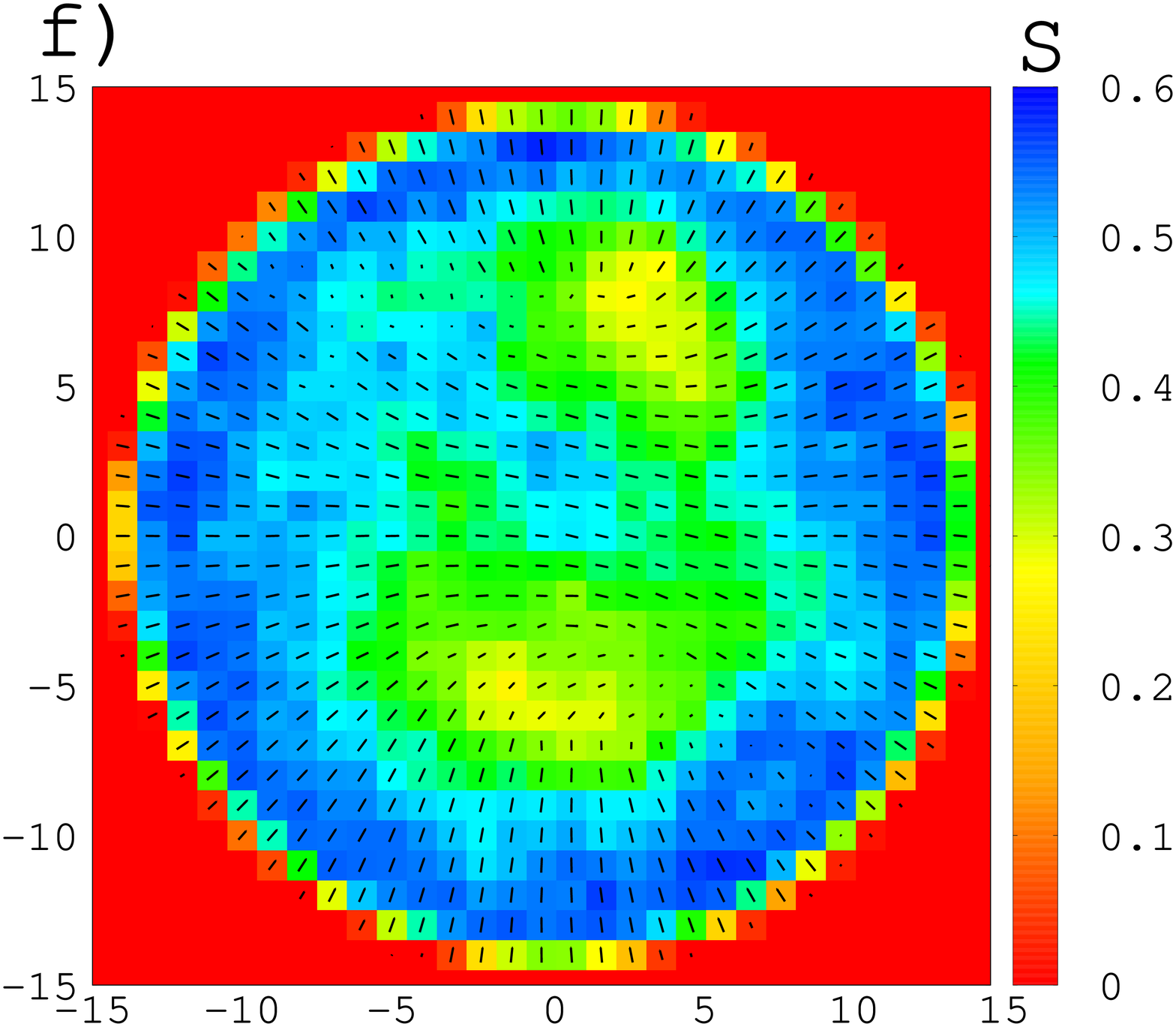}}}\vspace{-3.5cm}\\
\subfloat{%
\resizebox*{6cm}{!}{\includegraphics{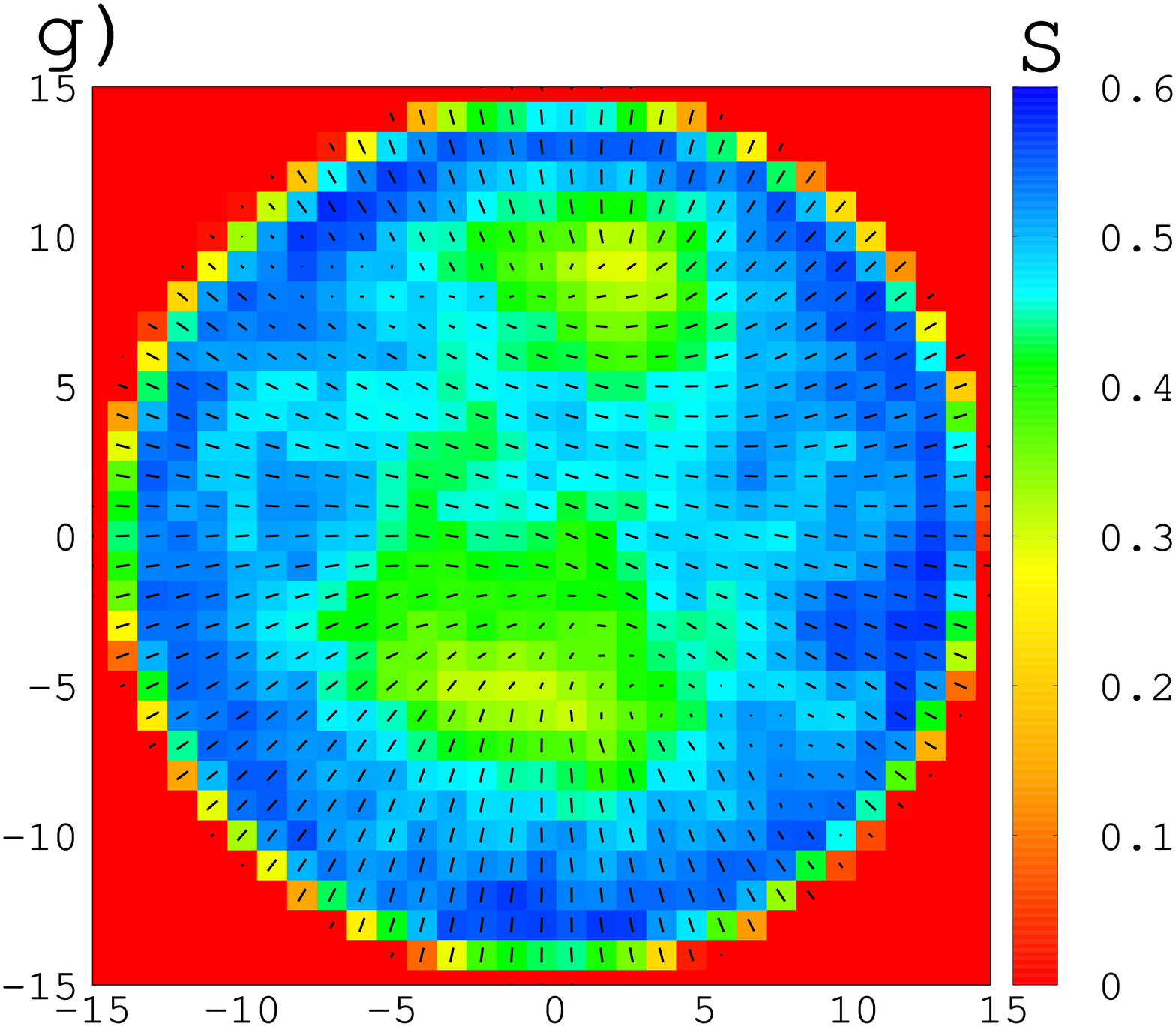}}}\vspace{-2cm}
\caption{Order parameter $S$ profile (colour map) and nematic director field (nail representation) of a cross section of the bridge for $a=0.6$ and $D=30$ at: a) $z=-12$, b) $z=12$, c) $z=-8$, d) $z=8$, e) $z=-4$, f) $z=4$, and g) $z=0$. Walls are located at $z=-15$ and $z=15$.} \label{fig5}
\end{figure}

For pore width $D=30$ the nanobridges have an aspect ratio $\Gamma \approx 1$. As in the previous case, barrel-shaped nanobridges are observed for $a=0.25$ and $a=0.6$, while for $a=1.0$ the nanobridge is hourglass-shaped. However, there is a dramatic change in the orientational order within the nanobridge for all the values of the wall-particle interaction strength $a$ we consider. For instance, an instantaneous snapshot of the nanobridge for $a=1.0$ is shown in Fig.~\ref{fig4}.  We can see that the orientational order is no longer axisymmetric around the vertical axis. Instead, in a large portion of the central part of the nanobridge, GB particles orient along a horizontal axis which we denote as $x'$. However, close to the walls particles orient vertically, while in two sides of the nanobridge, close to the nematic-vapour interface, particles orient along an horizontal direction $y'$ orthogonal to $x'$. The non-symmetrical character of the orientational order means that we have to use Eqs.~(\ref{defrhoz}) and (\ref{defqz}) to evaluate the density and orientational order parameter profiles. As a consequence, these profiles will be subject to stronger statistical uncertainties, since the size of voxels are smaller than the shells considered previously. Fig.~\ref{fig5} shows the scalar nematic order parameter $S$ and the nematic director orientational field of horizontal cross-sections of the nanobridge at different heights for $a=0.6$ (similar results are observed for $a=0.25$). We see that, close to the walls (i.e. $z=\pm 12$) the orientational field is consistent with an escaped radial configuration \cite{Vilfran1991,Crawford1992,Kralj1995,DeLuca2007}, in which the nematic director field deforms smoothly from a vertical orientation in the center to a homeotropic anchoring on the nematic-vapour interface. As we move towards the center, a low orientational order region emerges inside the nanobridge (in our case, for $z\approx \pm 8$). As the center of the nanobridge is approached, this low orientational order region splits in opposite sides with respect to the center of the nanobridge section, and the in-plane nematic director field is consistent with a polar planar texture \cite{Vilfran1991,Crawford1992,Kralj1995,DeLuca2007}, in which there are two $+1/2$ vertical disclination lines (see subplots Fig.~\ref{fig5}e), f) and g) for $z=\pm 4$ and $z=0$). This is a clear indication of the formation of a $+1/2$ disclination ring in a plane $y'z$, although our results show that its position is subject to strong fluctuations.
\begin{figure}
\centering
\subfloat{%
\resizebox*{6cm}{!}{\includegraphics{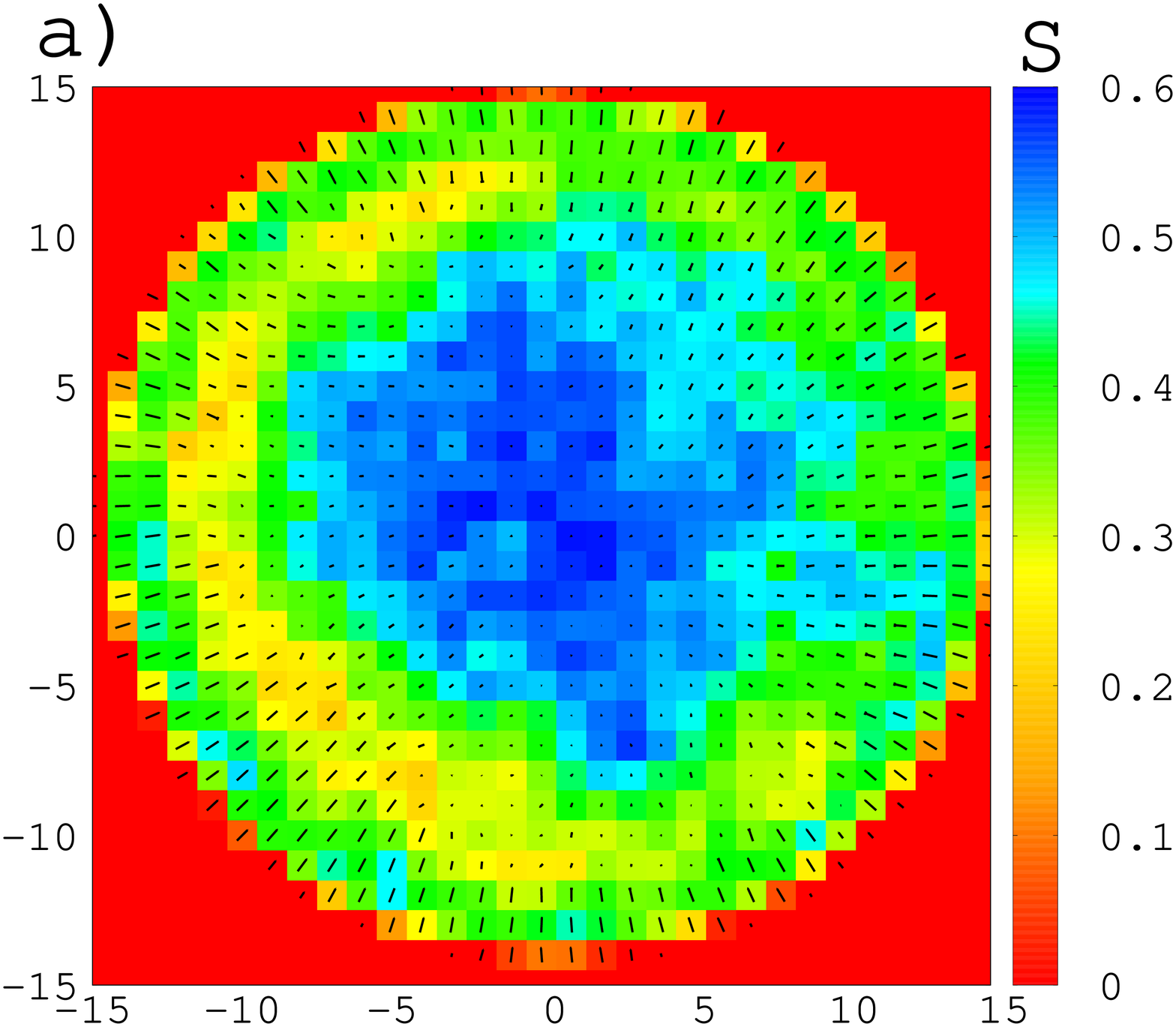}}}\hspace{5pt}
\subfloat{%
\resizebox*{6cm}{!}{\includegraphics{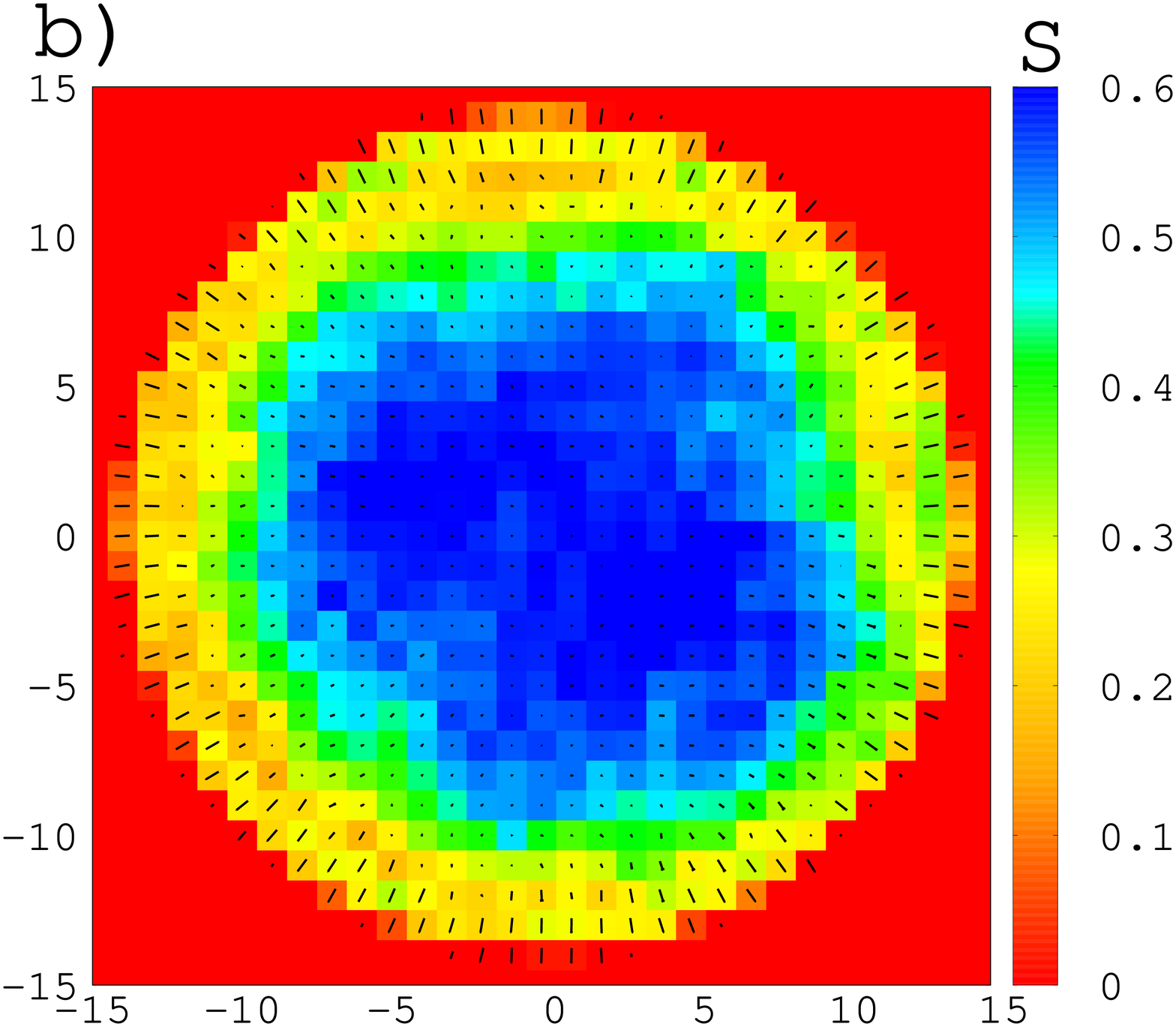}}}\vspace{-3.5cm}\\
\subfloat{%
\resizebox*{6cm}{!}{\includegraphics{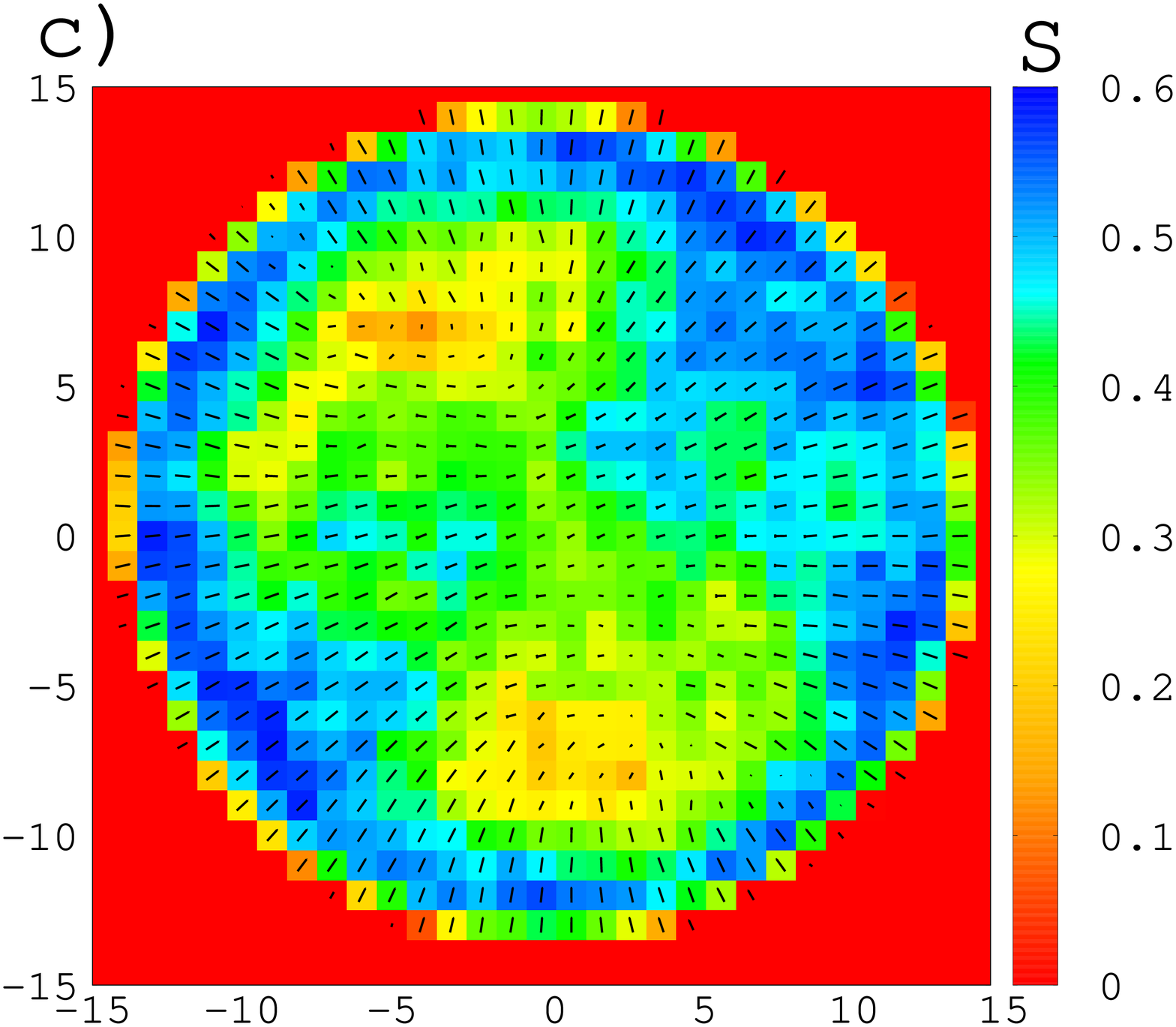}}}\hspace{5pt}
\subfloat{%
\resizebox*{6cm}{!}{\includegraphics{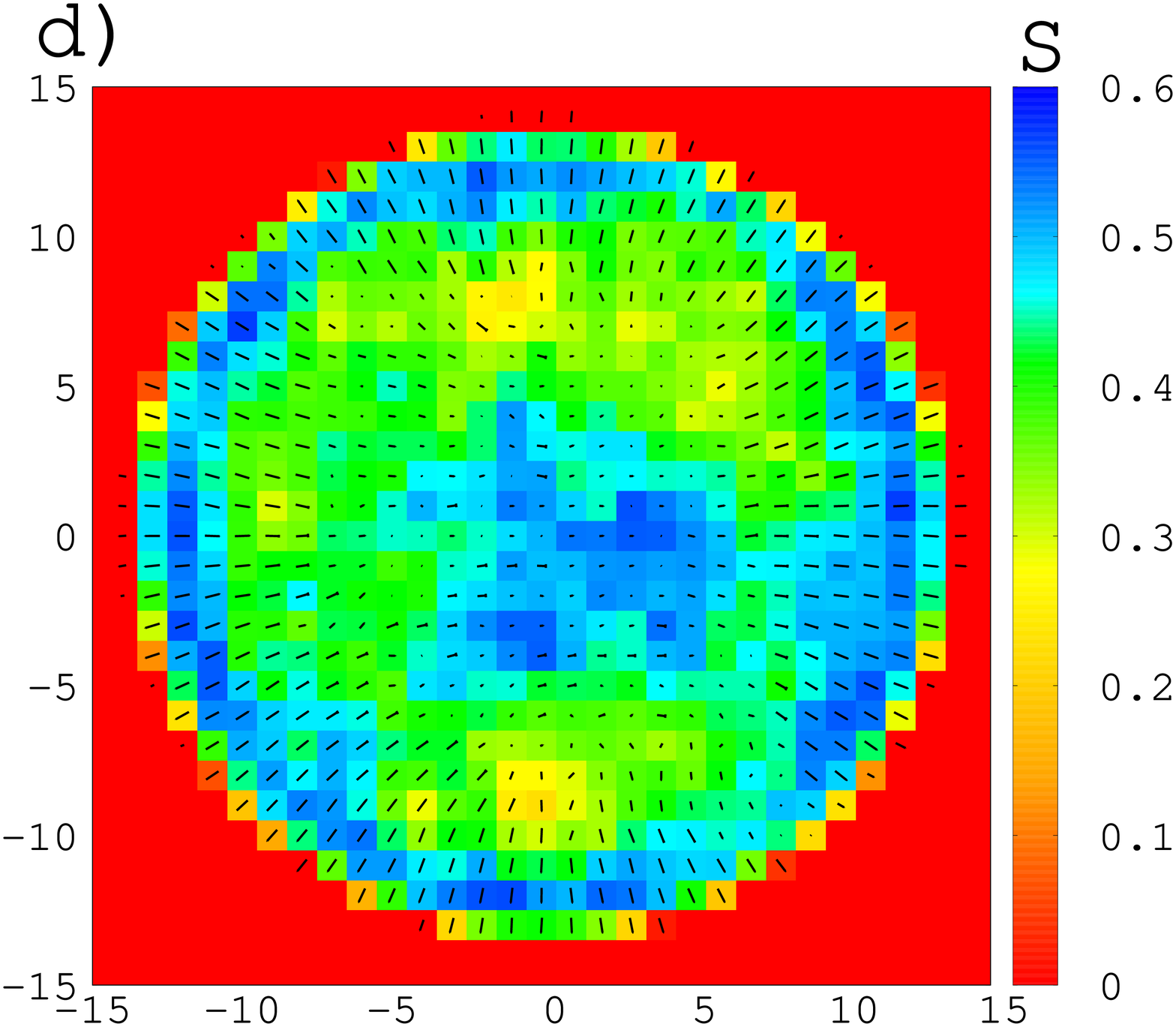}}}\vspace{-3.5cm}\\
\subfloat{%
\resizebox*{6cm}{!}{\includegraphics{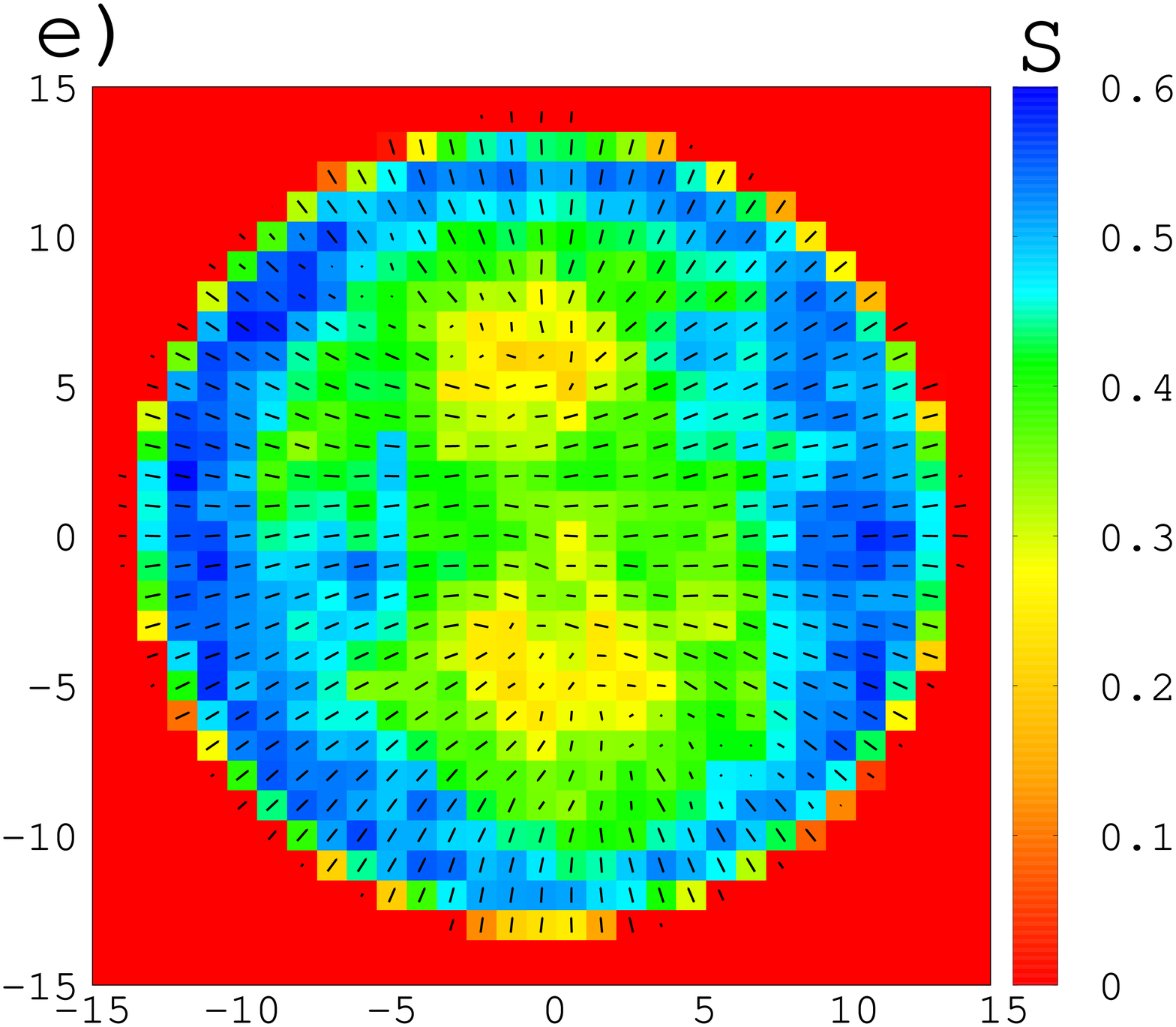}}}\hspace{5pt}
\subfloat{%
\resizebox*{6cm}{!}{\includegraphics{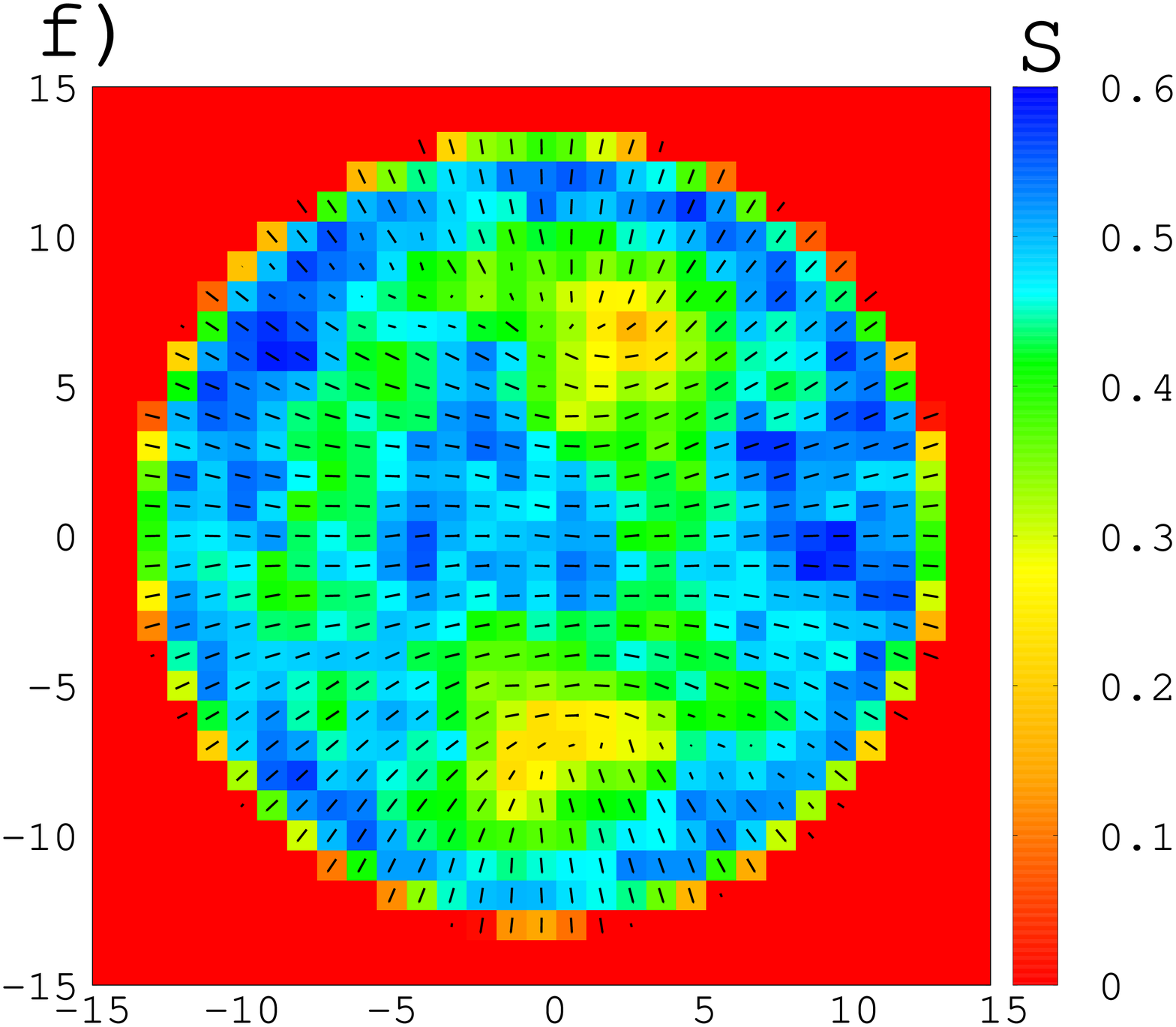}}}\vspace{-3.5cm}\\
\subfloat{%
\resizebox*{6cm}{!}{\includegraphics{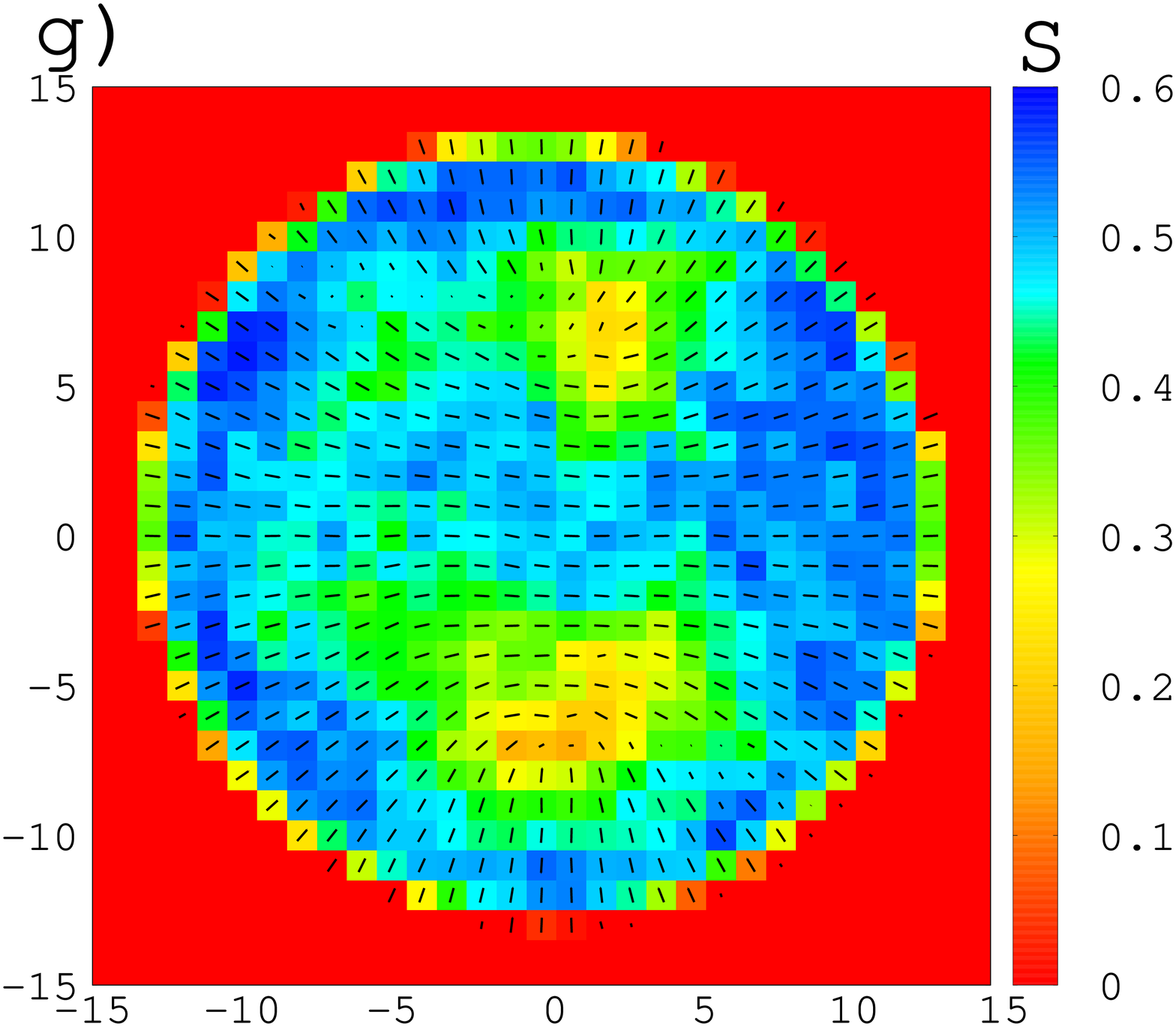}}}\vspace{-2cm}
\caption{Same as Fig.~\ref{fig5} for $a=1.0$ and $D=30$.} \label{fig6}
\end{figure}
For $a=1.0$, the qualitative picture is similar to the described one above. However, the picture is different close to the walls, as we can see in Fig.~\ref{fig6}. As we can see, for $z=\pm 12$ a ring of lower values of $S$ is formed close to the nematic-vapour interface. This structure may be related to the wall-nanobridge contact lines, as we saw above for $D=20$. As we approach the nanobridge center, two low orientational order regions emerge from the nematic-vapour interface (see Figs.~\ref{fig6}c) and d) for $z=\pm 8$), which move inside the nanobridge to form the $+1/2$ vertical disclination lines. Therefore, in this case there is no $+1/2$ disclination ring but a couple of disclination lines which emerge and disappear on the nematic-vapour interface.
\subsubsection{D=40}
\begin{figure}
\centering
\includegraphics{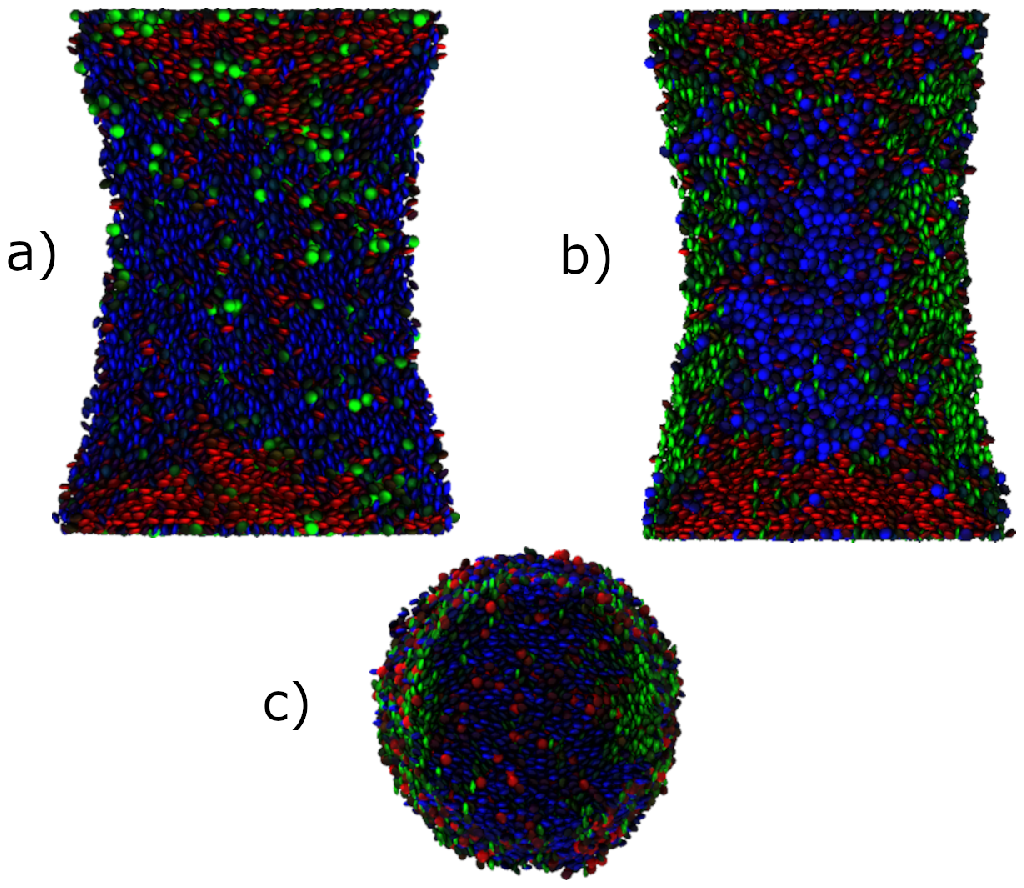}
\caption{Same as Fig.~\ref{fig4} for $D=40$ and $a=1.0$.} \label{fig7}
\end{figure}
\begin{figure}
\centering
\subfloat{%
\resizebox*{6cm}{!}{\includegraphics{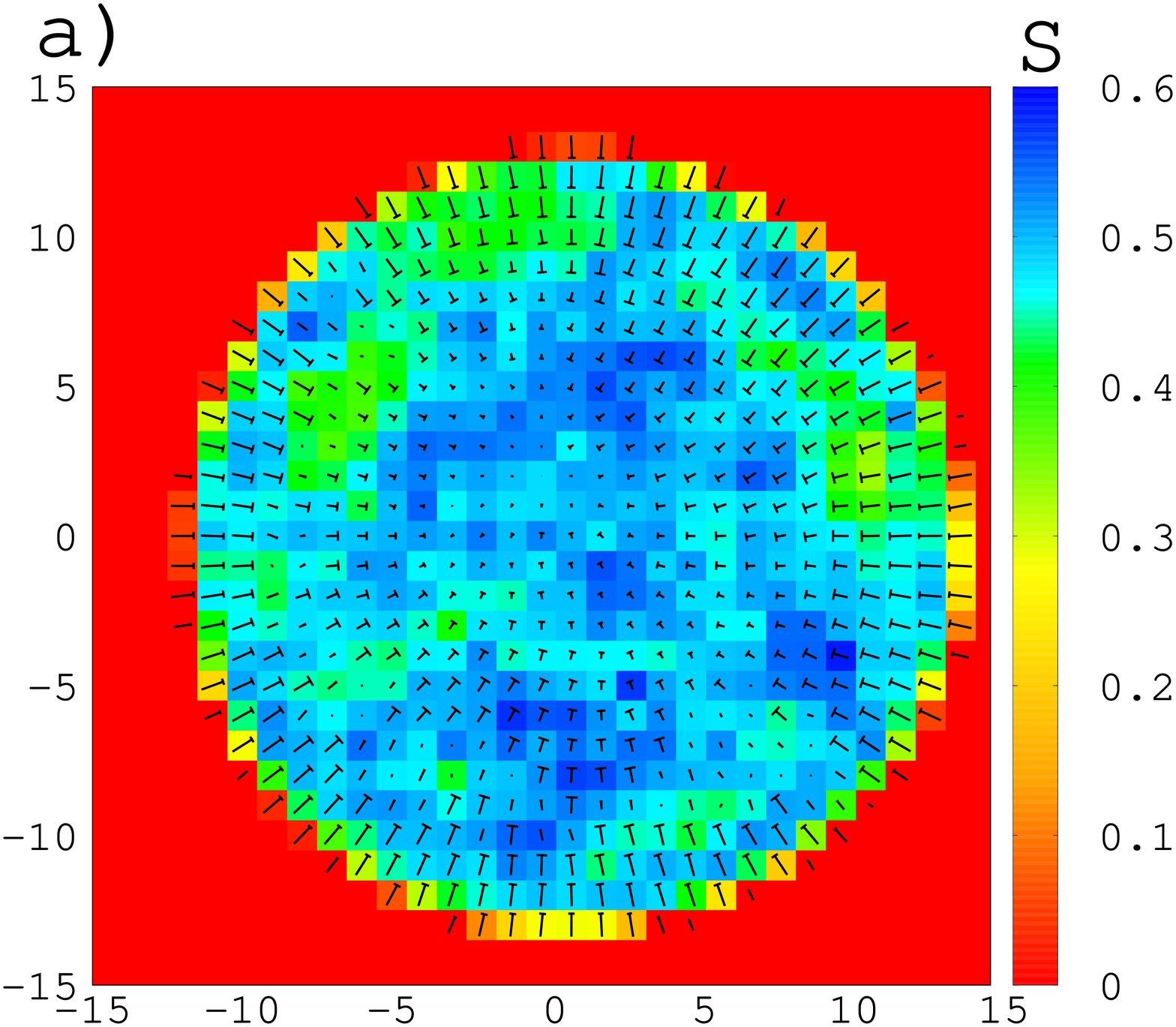}}}\hspace{5pt}
\subfloat{%
\resizebox*{6cm}{!}{\includegraphics{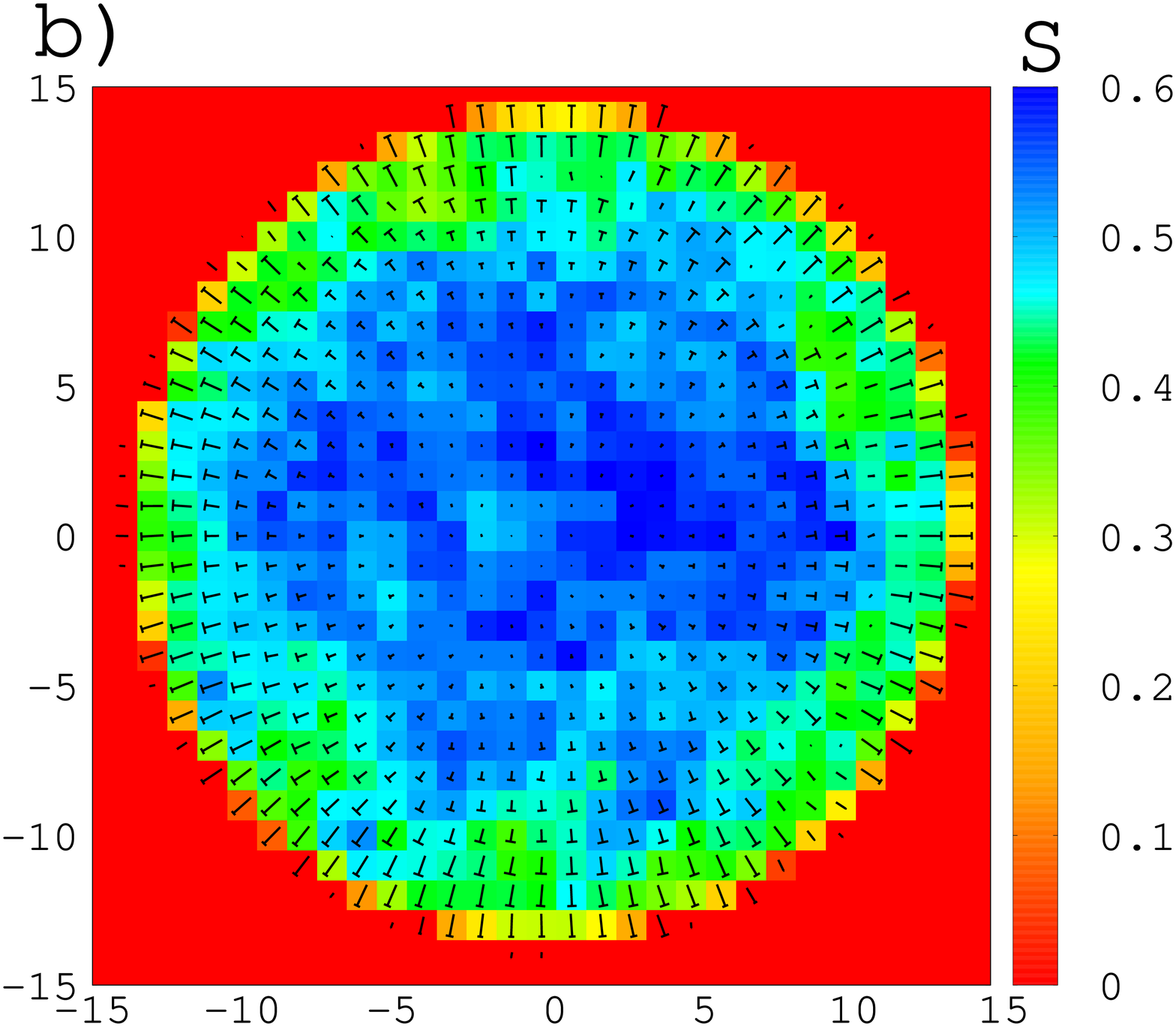}}}\vspace{-3.5cm}\\
\subfloat{%
\resizebox*{6cm}{!}{\includegraphics{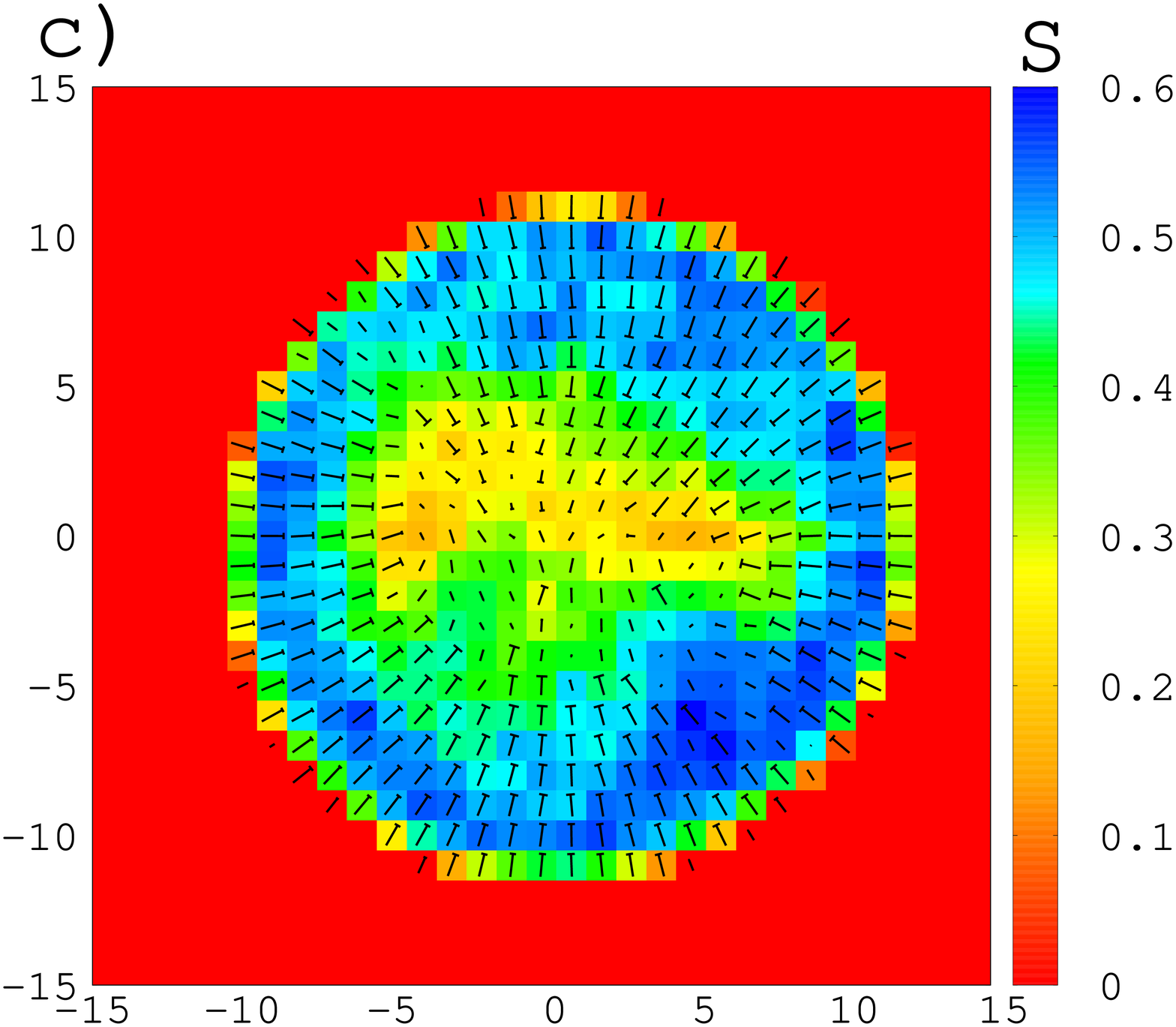}}}\hspace{5pt}
\subfloat{%
\resizebox*{6cm}{!}{\includegraphics{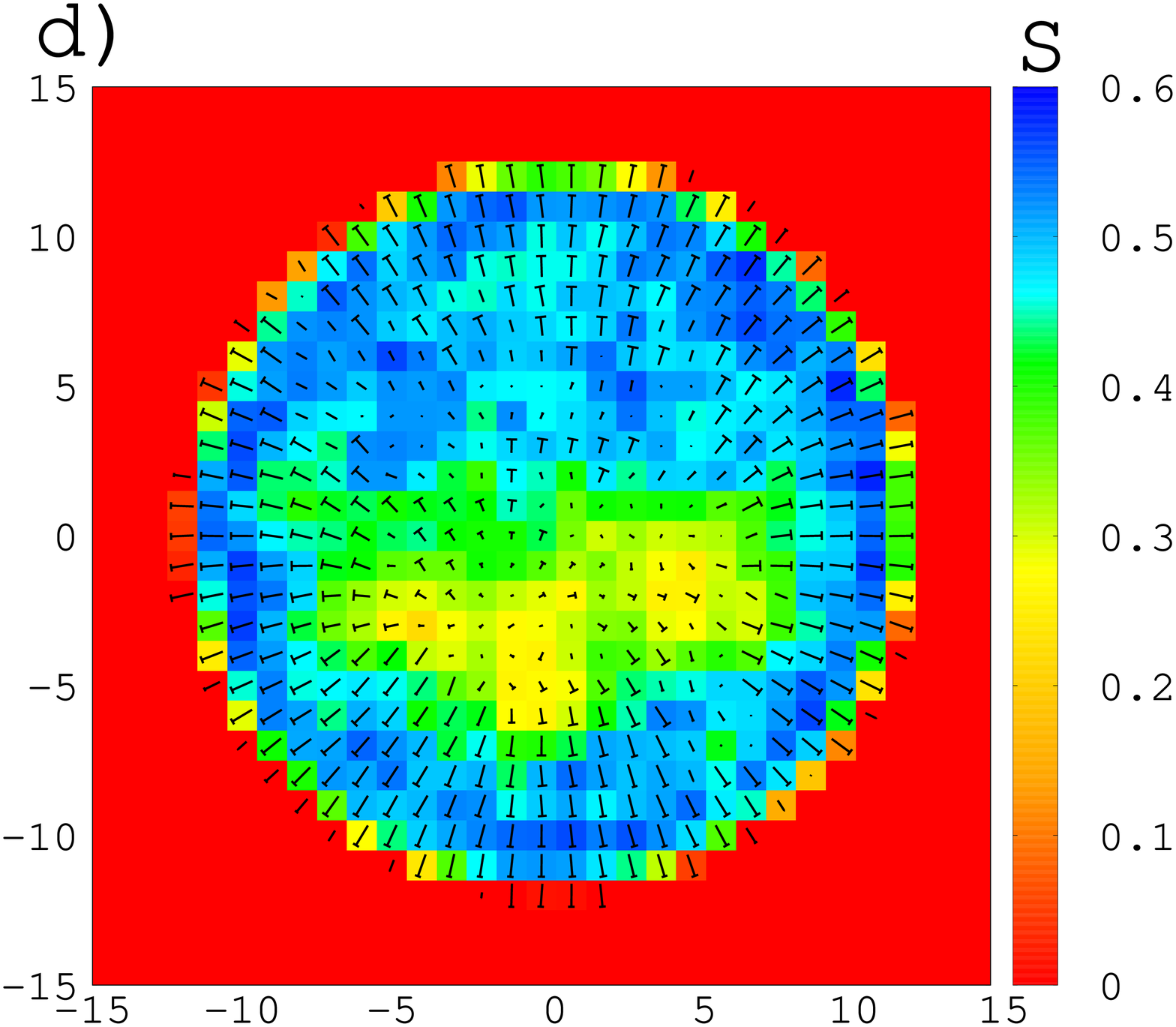}}}\vspace{-3.5cm}\\
\subfloat{%
\resizebox*{6cm}{!}{\includegraphics{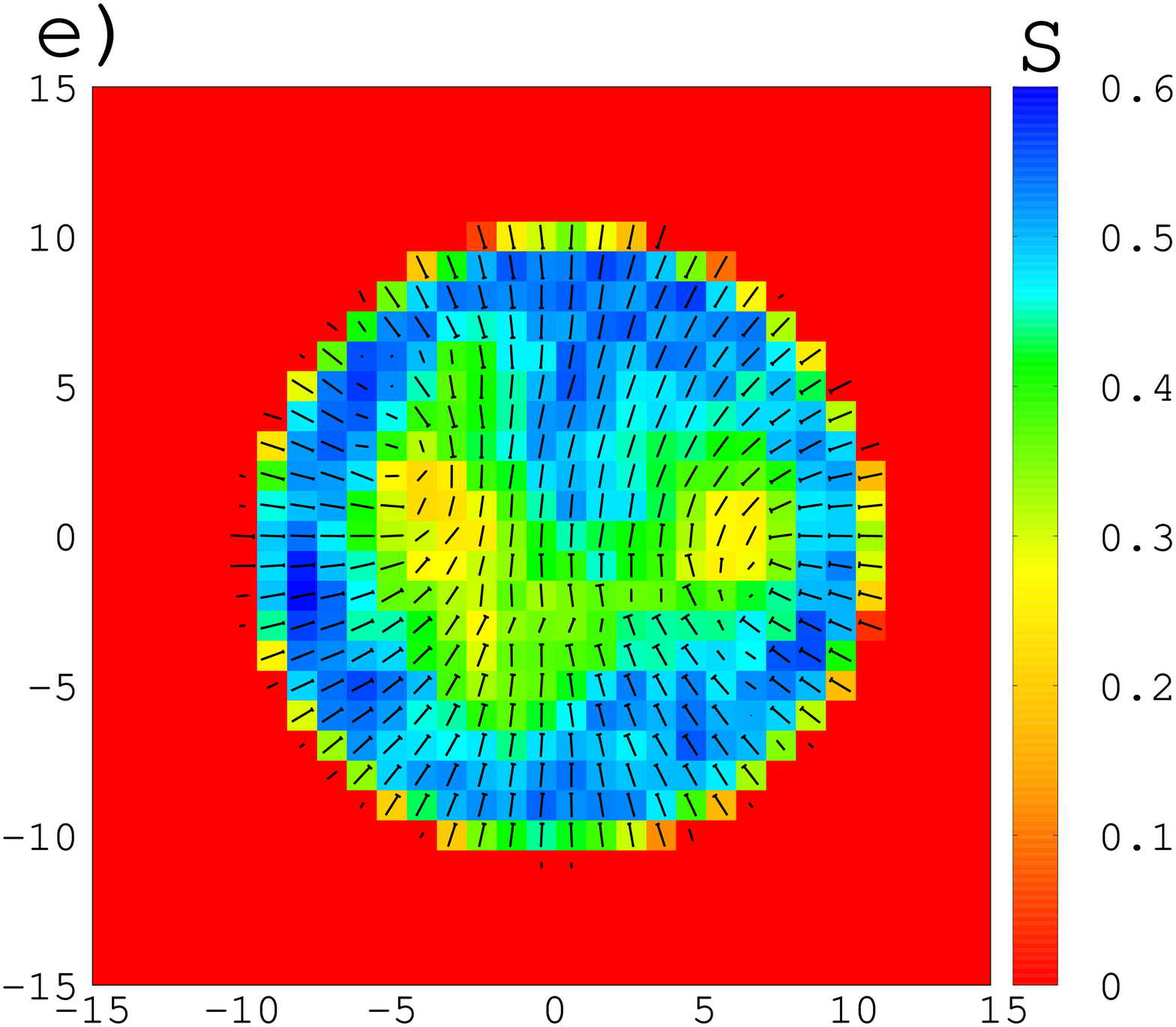}}}\hspace{5pt}
\subfloat{%
\resizebox*{6cm}{!}{\includegraphics{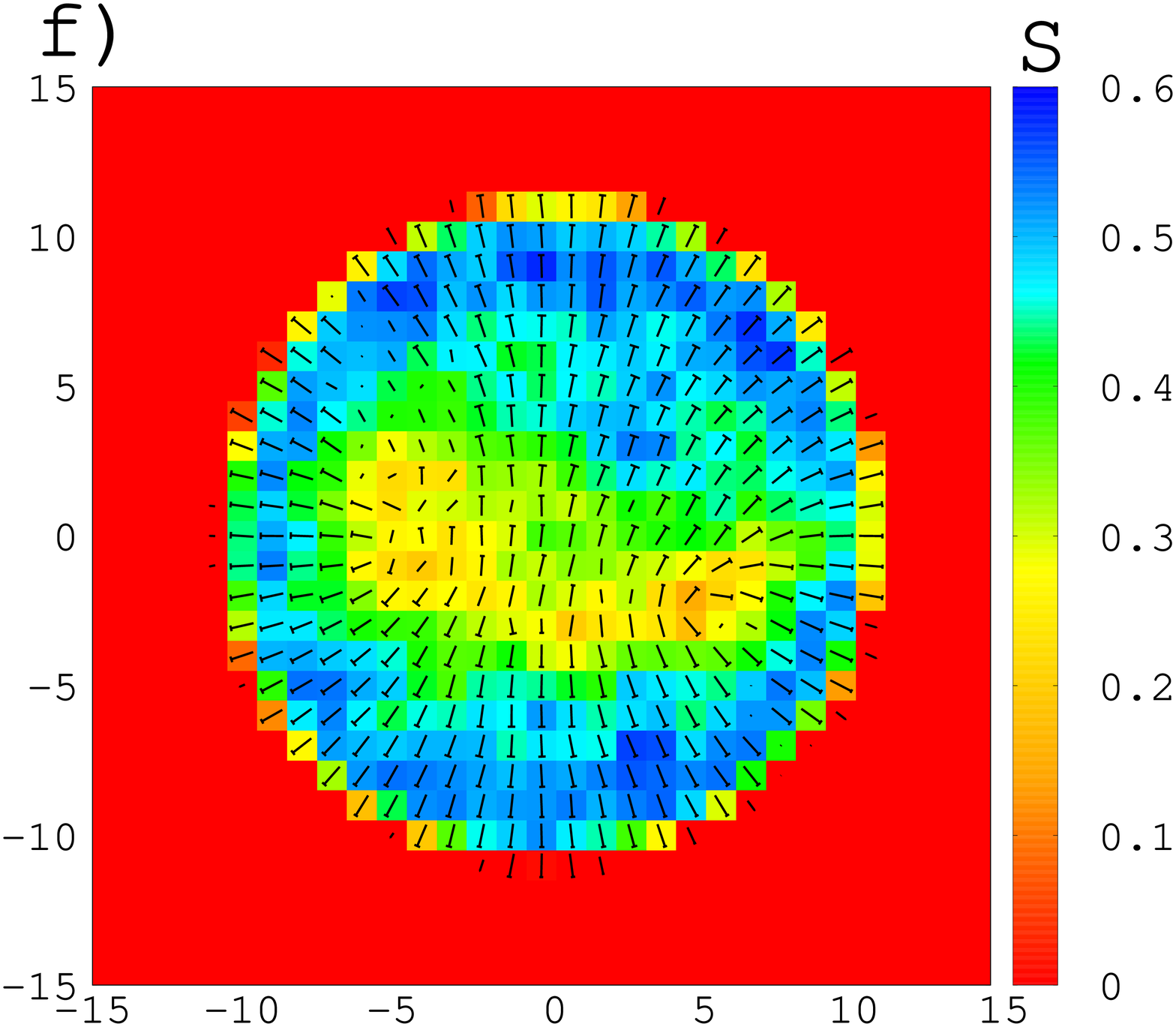}}}\vspace{-3.5cm}\\
\subfloat{%
\resizebox*{6cm}{!}{\includegraphics{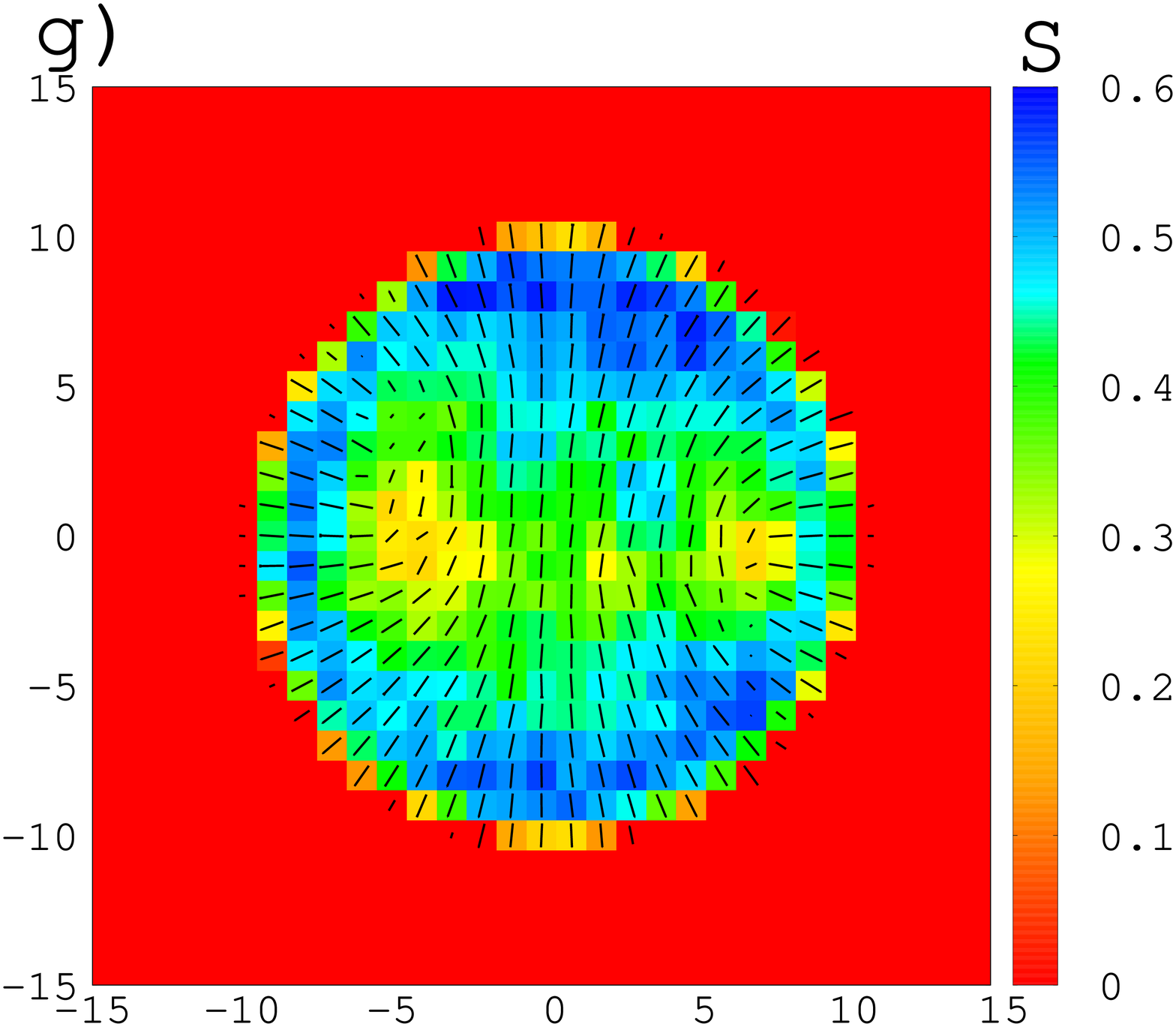}}}\vspace{-2cm}
\caption{Order parameter $S$ profile (colour map) and nematic director field (nail representation) of a cross section of the bridge for $a=1$ and $D=40$ at: a) $z=-16$, b) $z=16$, c) $z=-10$, d) $z=10$, e) $z=-6$, f) $z=6$, and g) $z=0$. Walls are located at $z=-20$ and $z=20$.} \label{fig8}
\end{figure}

Finally, we report our results for $D=40$. As mentioned above, we restrict ourselves the wall-particle interaction strength values $a=0.6$ and $a=1.0$, since for $a=0.25$ no nanobridge is formed inside the slit pore. In this case, the aspect ratio $\Gamma<1$, being the nanobridge barrel-shaped for $a=0.6$ and hourglass-shaped for $a=1.0$. Fig.~\ref{fig7} shows that the formation of a $+1/2$ disclination ring on a vertical plane, similar to the reported for $D=30$, is more clear as the nanobridge is stretched. This observation is confirmed by the scalar nematic order parameter profiles and nematic director fields on cross sections of the nanobridge (see Fig.~\ref{fig8}). For the latter, we used a nail representation, in which tilted particles are represented by nails, proportional in length to the vertical component of the nematic director. We see that these are more evident than in the $D=30$ case. The nematic texture is consistent with an escaped radial configuration close to the walls (i.e. $z=\pm 16$). At $z\approx \pm 6$, we observe the emergence of a low order region in the middle of the nanobridge, which splits into a couple of two opposite regions of low order as we approach the nanobridge middle plane. In the latter, the nematic texture is analogous to a a planar polar configuration in which two vertical $+1/2$ disclination rings are formed on opposite sides of the cross-section center. As we see, these results are analogous with those reported for the $a=0.6$, $D=30$ case. Now, most particles in the nanobridge orient along a horizontal axis, so do the global nematic director $\mathbf{N}$.  
\section{Discussion and conclusions}
In this paper we report computer simulation studies of nematic nanobridges between slit pores. The model that we consider shows homeotropic anchoring both on the walls and the nematic-vapour interface. This choice is motivated by the existence of experimental studies of 5CB capillary bridges on glass surfaces with homeotropic anchoring \cite{Gilli1997,Thiberge1999,Ellis2018}. However, we stress that our model does not try to match actual experiments, but we use a simple model which help us to understand the phenomenology these systems may present. We have considered three different values of pore width, in which the nanobridge has an aspect ratio $\Gamma >1$, $\Gamma \approx 1$ and $\Gamma < 1$. On the other hand, we also explored different values of the wall-particle interaction, which determine the wetting properties of the wall by the nematic and, consequently, we obtain either barrel-like or hourglass-like nanobridges. We observe that, for $\Gamma>1$, a nanobridge with a $+1/2$ disclination ring is formed on the nanobridge equatorial plane if the nanobridge is barrel-shaped. On the other hand, a similar structure in which an extended core which connects both contact lines between the walls and the nanobridge is observed for hourglass-shaped nanobridges. For $\Gamma \lesssim 1$, the nematic texture within the nanobridge changes dramatically, as now in general a $+1/2$ disclination ring is formed on a vertical plane. Only for $\Gamma \approx 1$ and hourglass-shaped nanobridges a slightly different texture is observed, as two vertical $+1/2$ disclination lines emerge and disappear on the nematic-vapour interface. 

Our findings have similarities and differences with respect to the experimental results reported in the literature \cite{Ellis2018}. As in the experiments, we see that the aspect ratio of the nanobridge is an essential property which determines the nematic texture of it. In particular, for narrow slit pores, i.e. $\Gamma>1$, and barrel-like nanobridges, we clearly observe an analogous structure to the shown in experiments, i.e. a radial or $+1/2$ disclination ring. For hourglass-like, $\Gamma>1$ capillary bridges, on the other hand, experiments report the formation of a hyperbolic or $-1/2$ disclination ring, which it is not the case in our simulations. Maybe smaller contact angles are needed to observe this case, or a larger nanobridge, see below. The main differences arise for the cases in which $\Gamma \lesssim 1$: instead of a texture characterized by a hyperbolic point defect located at the center of the bridge, our simulations show a texture with a vertical $+1/2$ disclination ring. 

In order to explain these discrepancies, we note that experiments consider bridges of sizes which spans over tenths to hundreds of micrometers, while our simulations are performed in a lengthscale of tenths of molecular diameters. In terms of nematic correlation lengths, which is of order of tenths of nanometers for 5CB and similar to the molecular diameter, experimental lengthscale is a factor of $10^3-10^4$ larger than our simulations. This fact may help to understand the discrepancies between experiments and our simulations. First, we note that no true point defect can be observed in our simulations, since simple calculations based on the Frank-Oseen functional show that, in reality, they are disclination rings of microscopic radius \cite{Terenjev1995}. Secondly, we can understand the discrepancies for $\Delta<1$ if we focus on the middle part of the bridge, which can be roughly approximated by a cylinder. Theoretical calculations show that, for infinitely long cylindrical capillaries and under conditions where no chiral states emerge, a planar-polar configuration is stable for narrow capillary radii, while an escaped radial configuration is observed for broad capillaries \cite{Vilfran1991,Crawford1992,Kralj1995,DeLuca2007}. Note that, in many circumstances, domains of opposite escaped radial configurations are separated by narrow disclination rings which can be assimilated to point defects \cite{DeLuca2007}. If we assume that these configurations deform continuously to the textures of the capillary bridges in their central region, the discrepancies would be consequence of the different lengthscales considered in experiments and computer simulations. 
So, the reproduction of the experimental results by computer simulation may need a much larger number of particles in the capillary bridge for a given aspect ratio, which could be unfeasible from computational point of view. However, it may be possible to test experimentally the simulation results by considering colloidal liquid crystals, such as suspensions of gibbsite platelets \cite{Verhoeff2010}. 

Finally, it would be interesting to locate the borderline between the capillary states corresponding to the low-$\Gamma$ (i.e. non-axisymmetric state) and large-$\Gamma$ (i.e. axisymmetric state) cases. Our results show that this transition occurs for a pore width between $20$ and $30$ for $N=32000$. Another aspect that we would like to address is the order of the phase transition. We expect this transition to be first-order since both states have different symmetry, but we cannot preclude more complex scenarios. These issues need a more systematic computer study of the capillary nanobridges which is beyond the scope of this paper. This analysis may be addressed in future works.

\section*{Acknowledgements}

This paper is dedicated to the memory of Prof. Luis F. Rull, the lead author of this paper who sadly passed away during its preparation. 

\section*{Disclosure statement}

The authors report there are no competing interests to declare.

\section*{Funding}

This work was supported by the Consejer\'{\i}a de Econom\'{\i}a, Conocimiento, Empresas y Universidad (Junta de Andaluc\'{\i}a, Spain) under Grants number US-1380729 and P20\_00816, cofunded by EU FEDER; and Ministerio de Ciencia e Innovaci\'on (Spain) under Grant number  PID2021-126348NB-I00.

\bibliographystyle{tfnlm}

\end{document}